\documentclass[preprint2]{aastex}

\usepackage{amsmath}
\usepackage[]{subfigure}
\usepackage{graphicx}

\allowdisplaybreaks
\shorttitle{}
\shortauthors{G\"urlebeck and Petroff}

\newcommand{\ba}{\bar{a}}
\newcommand{\bA}{\bar{A}}
\newcommand{\bB}{\bar{B}}
\newcommand{\bO}{\bar{\Omega}}
\def\APJ{Astrophys.\ J.}

\def\JMP{J.\ Math.\ Phys.}

\def\PRD{Phys.\ Rev.\ D}

\def\PTP{Prog.\ Theor.\ Phys.}
\def\PTRSL{Phil.\ Trans.\ R.\ Soc.\ Lond.\ A}

\graphicspath{{Figures/}}

\begin{document}

\title{A generalized family of Post-Newtonian Dedekind Ellipsoids}

\author{Norman G\"urlebeck}
\affil{ZARM, University of Bremen, Germany, EU}
\affil{Institute of Theoretical Physics, Charles University, Prague, Czech
Republic, EU}
\email{norman.guerlebeck@zarm.uni-bremen.de}

\author{David Petroff}
\affil{Coordination Centre for Clinical Trials, University of Leipzig, Leipzig,
Germany, EU}
\email{david.petroff@zks.uni-leipzig.de}

\begin{abstract}
We derive a family of post-Newtonian (PN) Dedekind ellipsoids to first order.
They describe non-axially symmetric, homogeneous, and rotating figures of
equilibrium. The sequence of the Newtonian Dedekind ellipsoids allows for an
axially symmetric limit in which a uniformly rotating Maclaurin spheroid is
recovered. However, the approach taken by \citet{Chandrasekhar_197478} to find
the PN Dedekind ellipsoids excludes such a limit. In
\cite{Gurlebeck_2010}, we considered an extension to their work that permits a
limit of 1 PN Maclaurin ellipsoids. Here we further detail the sequence
and demonstrate that a choice of parameters exists with which the singularity
formerly found in \cite{Chandrasekhar_197478} along the sequence of PN
Dedekind ellipsoids is removed.
\end{abstract}


\section{Introduction}\label{sec:introduction}

The modeling of equilibrium figures is one major goal in astrophysics. In the case
of compact objects, relativistic effects become relevant and it is necessary to treat them
using General Relativity. Exact solutions in this field are rare, and one is
generally required to resort to approximation or numerical methods. A detailed treatment
of these issues can be found in Meinel et al. (2008), Friedman \& Stergioulas (2013), where
the latter includes important discussions of stability. Although it is the numerical approach,
with its many possibilities for taking into account the properties of matter, that provides
the most accurate means of modeling real astrophysical objects, it is essential that one
pursues analytic methods in order to gain deeper insight into the nature of the theory of
relativity and how it differs from Newtonian theory. One of the most fruitful avenues of
research has always been the treatment of homogeneous matter, where it is precisely the
extreme simplification that has permitted great headway. Here we consider one such object,
the post-Newtonian (PN) Dedekind ellipsoid, in some depth, since we believe that doing so
provides new insights by correcting previous errors and by setting out along a path that
may provide new answers concerning the nature of stationary solutions within relativity.

In Newtonian gravity, closed form solutions are known for various ellipsoidal figures
of equilibrium including the Maclaurin spheroids (rigidly rotating, axially symmetric, and
stationary), the tri-axial Jacobi ellipsoids (rigidly rotating, non-axially symmetric, and
time independent in a rigidly rotating frame), the tri-axial Dedekind ellipsoids (non-axially
symmetric and stationary), and the tri-axial Riemann ellipsoids (in general: non-axially
symmetric and time independent in a rigidly rotating frame). For a summary of
their properties see \cite{Hagihara_1970,Chandrasekhar_1987}. The fluids
described by these solutions are homogeneous and have a velocity field, which is
linear in Cartesian coordinates. Moreover, they allow for the ratio of two half
axes of the ellipsoid to be adjusted arbitrarily giving rise to one parameter
families of solutions. For tri-axial ellipsoids, the third semi-axis is uniquely
determined by choosing this ratio. The Jacobi sequence and the Dedekind sequence
branch off from the Maclaurin sequence thereby allowing an axially symmetric and
rigidly rotating limit.

In the search for analogous figures of equilibrium in General Relativity, there
is still much work to do. However, for all of the aforementioned Newtonian
families of figures of equilibrium, a post-Newtonian (PN) approximation was
found in the sequence of papers
\cite{Chandrasekhar_1965b,Chandrasekhar_1967a,Chandrasekhar_1967c,
Chandrasekhar_1970a,Chandrasekhar_1971a,Chandrasekhar_1971b,Chandrasekhar_197478}.
Hereafter we call these articles Papers I and, if we refer only to the last one,
Paper II. The Maclaurin sequence was also studied in \cite{Bardeen_1971,
Petroff_2003a}. In the latter paper, an algorithm was given that allows one to
obtain the PN approximation to the Maclaurin ellipsoids to arbitrary order.
Since the Newtonian Jacobi ellipsoid has a time dependent quadrupole moment, the
energy loss due to gravitational wave emission can be estimated. This was done
in \cite{Chandrasekhar_1970a}. Thus, they cannot describe figures of equilibrium
in General Relativity, assuming that the Newtonian limit exists. In fact, it was
shown that the non-radiating final state is the Maclaurin ellipsoid at the
bifurcation point assuming that the Jacobi ellipsoids evolve along the Jacobi
sequence. Similarly, the Riemannian ellipsoids also lose energy due to
gravitational wave emission. The irrotational Riemann ellipsoids were
investigated to 1 PN order in \cite{Taniguchi_1998}. However, the Newtonian
Dedekind ellipsoids are stationary. Although they are non-axially symmetric they
keep their form due to internal motion -- each fluid element moves along an
ellipse in a plane perpendicular to the total angular momentum of the
configuration. Thus, they are a good starting point to investigate the question
if \emph{stationary but not static relativistic stars are necessarily axially
symmetric}, cf. \cite{Lindblom_1992}. If dissipative effects are not neglected,
it was shown in \cite{Lindblom_1976} that this is always true, but in case of
perfect fluids it is still an open question.

Before we start deriving PN Dedekind ellipsoids, we should state which
properties should be satisfied by such a sequence. Obviously, they should yield
the Newtonian Dedekind ellipsoids in an appropriate limit, which is guaranteed
if they are used as starting point of a PN approximation. Moreover, the PN
Dedekind ellipsoids\footnote{Of course, the shape of the 1 PN configuration is
not necessarily that of an ellipsoid in the coordinate system chosen. However,
since the corrections are small we still refer to the solution as an
``ellipsoid''.} should generalize as many properties of the Newtonian Dedekind
ellipsoids as possible. This includes the reflection symmetry with respect to
the coordinate planes and they should approach the 1 PN Maclaurin spheroids
close to the bifurcation point. The first property is respected in Paper II but
not the second. In \cite{Gurlebeck_2010}, we showed that both requirements can
be satisfied with a generalization of the ansatz for the 1 PN Dedekind
ellipsoids of Paper II. In the present paper, we study the entire sequence of
these figures of equilibrium and discuss their properties.

All 1 PN sequences of figures of equilibrium studied in Papers I admit
singularities at certain axis ratios. This implies that in a neighborhood of
these points, the 1 PN approximation is not applicable any longer.
Interestingly, these singularities along the PN Maclaurin and PN Jacobi
sequences appear at axis ratios, where the Newtonian sequence has special
properties. In \cite{Chandrasekhar_1967a}, it was shown that the singularity
along the 1 PN Maclaurin sequence coincides with the first bifurcation point
along the Newtonian Maclaurin sequence of a sequence of axially symmetric,
stationary, rigidly rotating and homogeneous figures of equilibrium. A
conjecture stating that all the bifurcation points of such sequences are
reflected by a singularity in the PN approximation, cf. \cite{Bardeen_1971}, was
proven in \cite{Petroff_2003a}.
There it was pointed out that, if the bifurcation points are ordered
appropriately, then there appears a singularity in the $n$th PN order for the
$n$th bifurcation point.

For the PN Jacobi sequence, see
\cite{Chandrasekhar_1967c}, the singularity at the 1 PN order is related to the
onset of a fourth-harmonic neutral mode of deformation of the Newtonian Jacobi
ellipsoids; also here the singularities in the PN approximation of Jacobi
ellipsoids is intimately tied to physical properties of the Newtonian
sequence. However, the axis ratio, where the singularity obtained for the 1 PN
Dedekind ellipsoids appears in Paper II, could not be identified with a special
point along the Newtonian sequence by the authors of that paper.
With our generalization of the ansatz for the Dedekind ellipsoids, see
\cite{Gurlebeck_2010}, we are able to show that this singularity can be removed
completely, suggesting that it is only due to an ansatz, which is too
restrictive. Surprisingly, Chandrasekhar \& Elbert already considered such a
generalization in Footnote 2 of Paper II, but discarded it because it was not
helpful in curing the singularity in their opinion.

The paper is organized as follows. We will first discuss the Newtonian solution.
On the one hand, this is done to fix our notation, but will also enable us to
motivate certain limits and properties for the 1 PN generalized Dedekind
ellipsoids. We will also determine the exterior solution in an explicit form
using ellipsoidal harmonics. In Section \ref{sec:1 PNapproach}, we give the
field equations for a 1 PN self-gravitating perfect fluid solution. These are solved
for the Newtonian Dedekind ellipsoids as starting point. Subsequently, we
discuss the properties of this 1 parameter family in detail, in particular the singularities
in parameter space. Some explicit formulae, lengthy calculations, and figures
are moved to the Appendices for readability.

\section{The Newtonian solution}\label{sec:Newtonian solution}

The Newtonian Dedekind ellipsoids describe rotating and tri-axial ellipsoids
with a homogeneous mass density. They are stationary in an inertial frame and
are solutions to the coupled Poisson and Euler equations:
\begin{align}\label{eq:Newtonian_field_equation}
  \Delta U=-4\pi G,\quad \nabla p=\mu \left(\nabla U-\left(\mathbf v\cdot
  \nabla\right)\mathbf v\right),
\end{align}
 where $U$ is the Newtonian gravitational potential, $\mu$ the homogeneous mass
 density (i.e., a constant) and $\mathbf v$ the Newtonian velocity of the
 particles. The surface of vanishing pressure, i.e., the surface of the
 configuration, is that of a tri-axial ellipsoid:
\begin{align}\label{eq:surface_Newtonian}
  S\left(x^1,x^2,x^3\right)=1-\sum\limits_{i=1}^3
  \left(\frac{x^i}{a_i}\right)^2=0.
\end{align}
We have chosen $a_1 \ge a_2 \ge a_3$ without loss of generality.
The three axes $a_i$ have to satisfy the relation
\begin{align}\label{eq:axis_ratio}
  a_1^2 a_2^2 A_{12}=a_3^2A_3.
\end{align}
The index symbols $A_{i_1\ldots}$ and $B_{i_1\ldots}$, $i_n \in \{1,2,3\}$, see
\cite{Chandrasekhar_1987}, are defined by
\begin{align}\label{eq:abbreviation index symbols}
\begin{split}
  z_i&=u+a_i^2,\quad H(u)=\frac{a_1 a_2 a_3}{\sqrt{z_1z_2z_3}},\\
  A_{i_1\ldots i_k}&=\int\limits_{0}^{\infty}H(u)\left(z_{i_1} \cdots
  z_{i_k}\right)^{-1} d u,\\
  B_{i_1 \ldots i_k}&=\int\limits_{0}^{\infty}H(u)u\left(z_{i_1}
  \cdots z_{i_k}\right)^{-1} d u.
\end{split}
\end{align}
It is also convenient to define
\begin{align}
  A_\emptyset&=\int\limits_{0}^{\infty} H(u) d u.
\end{align}

These index symbols satisfy various identities that allow one to express any
$A_{i_1\ldots}$ and $B_{i_1\ldots}$ as a linear combination of $A_1$ and $A_2$,
see \cite{Chandrasekhar_1987}. We introduce the dimensionless axis ratios
$\ba_2=\frac{a_2}{a_1}$ and $\ba_3=\frac{a_2}{a_1}$. The index symbols are
homogeneous functions in $a_1$:
\begin{align}
\begin{split}
  A_{i_1 \ldots i_k}(a_1,a_2,a_3)&=a_1^{2-2k} 
  A_{i_1 \ldots i_k}(1,\ba_2,\ba_3)\\
  B_{i_1 \ldots i_k}(a_1,a_2,a_3)&=a_1^{4-2k}
  B_{i_1 \ldots i_k}(1,\ba_2,\ba_3).
\end{split}
\end{align}
The dimensionless index symbols $A_{i_1 \ldots i_k}(1,\ba_2,\ba_3)$ and
$B_{i_1\ldots i_k}(1,\ba_2,\ba_3)$ will be denoted by $\bA_{i_1 \ldots i_k}$
and $\bB_{i_1\ldots i_k}$, respectively. Thus, Equation \eqref{eq:axis_ratio}
in dimensionless form is given by $\ba_2^2 \bA_{12}=\ba_3^2 \bA_3$. It can be
solved numerically and determines $\ba_3$ as a function of $\ba_2$
independently of $a_1$ (see \cite{Chandrasekhar_1987} for a table of values for
$\ba_3$ given $\ba_2$). Hence, the free parameters of this solution are
$\ba_2\in [0,1]$, $a_1$ and the mass density $\mu$ or, alternatively, the total
mass of the configuration $M=\tfrac 4 3\pi\mu a_1a_2a_3$. We further introduce
a dimensionless constant $\bO$ and a constant $\Omega$ via
\begin{align}
  \bO=\sqrt{2\bB_{12}}=\tfrac{\Omega}{\sqrt{\pi G\mu}}.
\end{align}

The solution of Equations \eqref{eq:Newtonian_field_equation} with a surface
given by Equations \eqref{eq:surface_Newtonian} and \eqref{eq:axis_ratio} reads in the
interior, cf. \cite{Chandrasekhar_1987},
\begin{align}\label{eq:Newtonian_solution}
\begin{split}
  \mathbf v& =\sqrt{\pi G \mu}\bO \left(- \tfrac{x^2}{\ba_2} ,\ba_2
  x^1,0\right),\\
  U&=2\pi G \mu\left(a_1^2
  \bA_{\emptyset}-\sum\limits_{i=1}^{3}\bA_i\left(x^i\right)^2\right),\\
  p&= \pi G \mu^2  a_3^2 \bA_3 \left(1 - \sum\limits_{i=1}^3 
  \left(\frac{x^i}{a_i}\right)^2\right).
\end{split}
\end{align}

\subsection{Limiting cases of the Dedekind sequence}\label{sec:limiting_cases}

Several limits are possible in the parameter space of this family of solutions.
Focusing on the axis ratio $\ba_2$, there are the limits $\ba_2\to 1$ and
$\ba_2\to 0$. In the latter case, we have $\ba_3\to 0$, too. In the first case,
the Dedekind ellipsoid approaches the Maclaurin ellipsoid with
$\ba^{M}_3=0.5827\ldots$ marking the well-known bifurcation point along the
Maclaurin sequence. At this point, both the Jacobi and the Dedekind ellipsoids
branch off. In this limit, the free parameters $a_1$ and the mass density $\mu$
can be prescribed as an arbitrary function of $\ba_2$ leading to several
qualitatively different possibilities depending on the behavior of these
functions. For instance, if $a_1\to 0$, which implies $a_2\to 0$ as well as
$a_3\to 0$, and if  moreover $M\to M_0<\infty$, then the Maclaurin ellipsoids
contract to a point particle with mass $M_0$. In this limit, the velocity fields
vanish as well. If on the other hand, $a_1\to\infty$, which implies $a_2,~a_3\to
\infty$, and $\mu$ approaches some value $0<\mu_0<\infty$ the entire space is
filled with a rigidly rotating perfect fluid. Of course, this solution becomes
unphysical for radii, where the fluid elements have an orbital velocity greater
than the velocity of light. In the limit, where both, $a_1$ and $\mu$, approach
some finite and positive values $a_{1,0}$ and $\mu_0$, a Maclaurin ellipsoid is
obtained. Here as well, $a_{1,0}$ is restricted by the physical requirement of
subluminal motion.

Let us turn our attention to $\ba_2\to 0$. In order to discuss this limit, we
shall need to use the following expansions\footnote{We make use of the common
Landau notation, where for two functions $f(x)$ and $g(x)$ we have $f\in
o_x(g)$, if $\lim\limits_{x\to 0} \frac{f(x)}{g(x)}=0$. If the dependent
variable is clear from context we drop the index of the Landau symbol.}
\begin{align}\label{eq:explicit_expansions}
\begin{split}
  \ba_2=&\bar{a}_3+\bar{a}_3^3 \left(\ln  4-3-2 \ln
  \bar{a}_3\right)+o(\bar{a}_3^4),\\
  \bA_1=&-2 \bar{a}_3^2 \left(\ln \frac{\bar{a}_3}{2}+ 1\right)+ \bar{a}_3^4
  \left(\vphantom{+\frac{13}{2}}\ln \bar{a}_3 \left(4 \ln
  \bar{a}_3+\right.\right.\\
  &\left.\left.9-8 \ln 2\right)+\frac{13}{2}+\ln 2 (\ln 
  16-9)\right)+o(\ba_3^5),\\
  \bA_2=&1+\bar{a}_3^2 \left(2 \ln \bar{a}_3+\frac{5}{2}-\ln
  4\right)+o(\bar{a}_3^3).
\end{split}
\end{align}
The expansions of the other index symbols can be inferred from the recursion
relations given in \cite{Chandrasekhar_1987}, cf.\ the comment after Equation
\eqref{eq:abbreviation index symbols}.

In the most interesting case, where $a_1\to a_{1,0}$ and $M\to M_0$ with
$0<a_{1,0}<\infty$ and $0<M_0<\infty$ implying $a_2\to 0$ and $a_3\to 0$, the
ellipsoids degenerate to a rod. Then the velocity field of Equation
\eqref{eq:interior solution Newtonian} always
diverges logarithmically, which follows from Equations
\eqref{eq:explicit_expansions}. In fact, two anti-parallel, non-interacting
streams in the $x^1$-direction with infinite velocity emerge
such that the solution is static. The same holds true for the case
$a_{1,0}=\infty$. In these cases, the global solution is not admissible for a PN
approximation sufficiently close to the limit $\ba_2=0$.  However, if $M_0=0$ is
approached sufficiently fast, the limit describes a Newtonian solution that can
be interpreted as two anti-parallel streams of massless particles with a finite
velocity. Hence, a PN approximation might be possible. We present the details of
this in Appendix \ref{sec:limit_rod}.

If $a_{1,0}=0$ and $0<M_0<\infty$, we obtain again a point mass. If the limit of
$-a_1 \ln \ba_3$ is sufficiently small for $\ba_2\to 0$, then the velocity is
subluminal during the limiting process.

\subsection{The exterior solution}\label{sec:exterior_solution}

We describe here the formalism with which the exterior solution of the 1 PN
equilibrium figures is obtained in closed form in Section \ref{sec:1
PNapproach}. As a practical example, we apply the algorithm to the Newtonian
Dedekind ellipsoids. We introduce ellipsoidal coordinates $\lambda^i$ with
$\lambda^1>k>\lambda^2>h>\lambda^3>0$ with $h^2=a_1^2-a_2^2$ and
$k^2=a_1^2-a_3^2$, see e.g.\ \cite{Byerly:1893}:
\begin{align}\label{eq:coordinatetransformation}
\begin{split}
  \left(x^1\right)^2&=\frac{\left(\lambda^1\right)^2 \left(\lambda^2\right)^2
  \left(\lambda^3\right)^2}{h^2k^2},\\
  \left(x^2\right)^2&=-\frac{1}{h^2\left(k^2-h^2\right)} \prod\limits_{i=1}^3
  \left(\left(\lambda^i\right)^2-h^2\right),\\
  \left(x^3\right)^2&=\frac{1}{h^2\left(k^2-h^2\right)} \prod\limits_{i=1}^3
  \left(\left(\lambda^i\right)^2-k^2\right).
\end{split}  
\end{align}
These coordinates cover the octant $x^i>0$, which is sufficient since the
problem is reflection-sym\-metric with respect to the surfaces $x^i=0$. We
assume that the PN configuration has this symmetry as well. The surface of the
ellipsoid is characterized by $\lambda^1=a_1$. Using these coordinates, a
separation of variables in the Poisson equation with a density $g$ the support
of which is an ellipsoid is possible. The solution $f$ of $\Delta f=-4\pi g$ will
be of the form
\begin{align}\label{eq:general solution Poissonequation}
\begin{split}
  \sum\limits_{n=0}^{\infty}\sum\limits_{m=1}^{2n+1} f_{m}^{n}\left(\lambda^1
  \right)E_m^n\left(\lambda^2\right)E_m^n\left(\lambda^3\right),
\end{split}
\end{align}
where the functions $E_m^n$ are the Lam\'e functions of the first kind. Their
definition and the first few members of this complete set of functions can be
found in \cite{Byerly:1893}. The function $f_m^n$ will be obtained as a solution
of the inhomogeneous Lam\'e equation
$L^n_m(\lambda^1)[f]=\tilde g_m^n\left(\lambda^1\right)$, where the Lam\'e
operator $L^n_m$ is given by
\begin{align}
\begin{split}
  L^n_m&(x)=\left(x^2-h^2\right)\left(x^2-k^2\right)\frac{d^2}{d x^2}+\\
  &x\left(2x^2-h^2-k^2\right)\frac{d}{d
  x}+\left(K_m^n-n\left(n+1\right)x^2\right).
\end{split}
\end{align} 
The characteristic values $K_m^n$ of the Lam\'e functions are also defined in
\cite{Byerly:1893}. The $\tilde g_m^n\left(\lambda^1\right)$ are the expansion
coefficients of the density $\mu$ with respect to the ellipsoidal harmonics:
\begin{align}
\begin{split}
  &\left(\left(\lambda^1\right)^2-\left(\lambda^2\right)^2\right)
  \left(\left(\lambda^1\right)^2-\left(\lambda^3\right)^2\right)g
  \left(\lambda^i\right)\\
  &=\sum\limits_{n=0}^{\infty}\sum\limits_{m=1}^{2n+1}  \tilde g_m^n
\left(\lambda^1\right)E_m^n\left(\lambda^2\right)E_m^n\left(\lambda^3\right).
\end{split}
\end{align}
Since $g$ vanishes outside of the ellipsoid, the equation becomes homogeneous
and the sole solution with the correct asymptotics is given by
\begin{align}
 f_m^n\left(\lambda^1\right)=C_m^n F_m^n\left(\lambda^1\right),
\end{align}
where $F_m^n$ denotes the Lam\'e functions of the second kind, see again
\cite{Byerly:1893}. These follow from the Lam\'e functions of the first kind via
\begin{align}\label{eq:Lame_Fuction_Second_Kind}
\begin{split}
  &F_m^n\left(x\right)= E_m^n\left(x\right)\times \\
  &\int\limits_x^\infty
  \left(E_m^n\left(u\right)\right)^{-2} \left((u^2 - h^2\right) \left(u^2 - 
  k^2\right))^{-\tfrac12} du.
\end{split}
\end{align}
To the orders, which appear in the present paper, the $F^n_m$ can be given
explicitly in closed form in terms of elliptic functions.

In general, we have to solve the inhomogeneous Lam\'e equation. However, we can
rely for all potentials that we have to calculate on a result by
\cite{Ferrers:1877}. There the interior solution of the Poisson equation with a
density that is polynomial in Cartesian coordinates inside an ellipsoid is
given. It is also established that this solution can be connected to an exterior
solution, which vanishes at infinity, such that the solution is
continuously differentiable everywhere. Thus, we obtain $f_m^n\left(\lambda^1\right)$ for
$\lambda^1<a_1$ simply by a coordinate transformation of Ferrers' interior
solution from Cartesian to ellipsoidal coordinates. Since the interior solution
is polynomial in Cartesian coordinates, the expansion of this in ellipsoidal
surface harmonics $E_m^n\left(\lambda^2\right)E_m^n\left(\lambda^3\right)$
terminates at finite order and we can simply read the $f_m^n$ off. Hence,
$C_m^n$ can afterwards be obtained by
\begin{align}\label{eq:determinining_Cmn}
  C_m^n=  f_m^n\left(a_1\right)F_m^n\left(a_1\right)^{-1}
\end{align}
and the potential $U$ is completely determined.

We illustrate this method for the Dedekind ellipsoids, which are of the type
considered by \cite{Ferrers:1877}. The interior solution can be written after a
transformation to ellipsoidal coordinates as
\begin{align}\label{eq:interior solution Newtonian}
\begin{split}
  U=&f_1^0\left(\lambda^1\right) E_1^0\left(\lambda^2\right)E_1^0
  \left(\lambda^3\right)+\\
  &\sum\limits_{i=1}^2f_i^2\left(\lambda^1\right) E_i^2\left(\lambda^2\right)
  E_i^2\left(\lambda^3\right),\\
  f_1^0\left(\lambda^1\right)=&-\frac{4}{3} \pi  G\mu \left(3
 \left(\lambda^1\right)^4-\right.\\
  &\,\,\,\left.2 \left(\lambda^1\right)^2 \left(h^2+k^2\right)+h^2
  k^2\right),
\end{split}
\end{align}
\begin{align*}
  f_{1/2}^2\left(\lambda^1\right)=&-2 \pi  G\mu \left(1\pm\frac{3
  \left(\lambda^1\right)^2-h^2-k^2}{\sqrt{h^4-h^2 k^2+k^4}}\right).
\end{align*}
This implies that the exterior solution has the form
\begin{align}
\begin{split}
  U=&C_1^0 F_1^0\left(\lambda^1\right)E_1^0\left(\lambda^2\right)
  E_1^0\left(\lambda^3\right) +\\
  &\sum\limits_{i=1}^2 C_i^2 F_i^2\left(\lambda^1\right)
  E_i^2\left(\lambda^2\right)E_i^2\left(\lambda^3\right),
\end{split}
\end{align}
where the constants $C_m^n$ are obtained from Equations
\eqref{eq:determinining_Cmn} and \eqref{eq:interior solution Newtonian}.

In fact, this procedure allows us to obtain the exterior solution for mass densities of the form $\mu_{ijk}=\mathrm{const.}\, (x^1)^i (x^2)^j (x^3)^k$ in closed form. This enables
us to determine the 1 PN metric in the exterior region in closed form for the
Dedekind ellipsoids as well as the Jacobi ellipsoids in exactly the same way.
Higher order ellipsoidal harmonics will be necessary in this scheme. However, the
calculations are tedious and will not be presented in detail here. We will only
repeat the form of the higher moments in the interior of the ellipsoid, which
are the starting point for the straightforward calculations, in Appendix
\ref{sec:higher_moments}.

\section{The 1 PN approach}\label{sec:1 PNapproach}

In \cite{Chandrasekhar_1965a}, a set of field equations was discussed whose
solutions describe perfect fluids dynamically to first order in $\tfrac{1}{c^2}$
in general relativity. In a subsequent series of papers
\citep{Chandrasekhar_1965b, Chandrasekhar_1967a, Chandrasekhar_1967c,
Chandrasekhar_1971a, Chandrasekhar_1971b, Chandrasekhar_197478}, solutions to
these equations were constructed using different Newtonian configuration as a
starting point-- namely Maclaurin ellipsoids, Jacobi ellipsoids and Dedekind
ellipsoids. Although these are all equilibrium figures, the field equations in
\cite{Chandrasekhar_1965a} allow for non-stationary solutions, too. Thus, one
always has to determine and to solve the equations belonging to
the dynamical aspects of the fluid. In contrast, the projection formalism
described in \cite{Geroch_1971} and the field equations derived therein
implement the stationarity from the beginning and can be used only in the
description of equilibrium figures. In our case, where we are primarily
interested in stationary solutions the latter equations are more advantageous
especially if one goes to higher PN orders. However, to the 1 PN order both
approaches yield the same result in the case of stationarity; discrepancies will
become apparent only at higher orders. For equations describing higher order PN
corrections, see, e.g., \cite{Chandrasekhar_1969a,Asada_1996,Asada_1996b}.

We use the same expansion of the metric as in \cite{Chandrasekhar_1965a}:
\begin{align}\label{eq:expansion of the metric}
\begin{split}
  g_{\alpha \beta}&=-\left(1+\frac{2U}{c^2}\right)\delta_{\alpha \beta},\\
  g_{\alpha 0}&=4U^{(3)}_{\alpha} c^{-3},\\
  g_{00}&=1-2U c^{-2}+2\left(U^{2}-\delta U-2 \Phi\right)c^{-4},
\end{split}
\end{align}
where Greek indices run from $1$ to $3$, $\delta_{\alpha \beta}$ denotes the
Kronecker delta, $x^0=c t$. $U$ is the Newtonian gravitational potential, which
we assume here to be that of a Dedekind ellipsoid as discussed in Section
\ref{sec:Newtonian solution}. The contribution $\delta U$ is defined
momentarily. 

The shape of the PN configuration is no longer that of the Newtonian ellipsoid
$S^{(0)}$ (cf.\ Equation \eqref{eq:surface_Newtonian}) and we denote it by
\begin{align}\label{eq:surfacePN}
  S = S^{(0)} + S^{(2)} c^{-2}.
\end{align}
Let us introduce a potential $U'$ that is a solution of the Poisson equation
\eqref{eq:Newtonian_field_equation} for this perturbed ellipsoid, i.e.,
\begin{align}\label{eq:PoissonUprime}
  \Delta U' = -4\pi G\mu\
\end{align}
with $\mu=\mathrm{const.}$ for $S<0$. The symbol $\delta U$ in Equation
\eqref{eq:expansion of the metric} is then defined by
\begin{align}\label{eq:U_prime}
  \delta U = (U'- U)c^2.
\end{align}
Note that we use the expansion parameter $c^{-1}$ to retain the compatibility with
Papers I and II. However, one could transform the results easily to a more
physical expansion parameter, e.g., $\varepsilon^2=\tfrac{2M G}{c^2 a_1}$.

The pressure and the velocity field are also expanded
\begin{align}
  p=p^{(0)}+p^{(2)}c^{-2},\quad v^\alpha=v^{(0)\alpha}+v^{(2)\alpha} c^{-2},
\end{align}
where $v^{(0)\alpha}$ is the Newtonian velocity\footnote{The three-velocity
$v^\alpha$ is defined as in Paper II and does not refer to the spatial
components of the four-velocity $u^i=\tfrac{d x^i}{d\tau}$, but is instead
defined as $v^\alpha=\tfrac{d x^\alpha}{d t}= c \tfrac{u^\alpha}{u^0}$.} and
$p^{(0)}$ is the Newtonian pressure, cf.\ Equations
\eqref{eq:Newtonian_solution}. We assume that the homogeneous mass density does
not change to any PN order.

The 1 PN equations of a self-gravitating perfect fluid read\footnote{Indices are
raised and lowered in these equations with the flat Euclidean metric.}, see
\cite{Chandrasekhar_1965a},
\begin{subequations}\label{eq:1 PN_equations}
\begin{align}
  \Delta \Phi &= -4\pi G\mu \left({v^{(0)}}^2+\frac{3p^{(0)}}{2
  \mu}+U\right),\label{eq:1 PNPhi}\\
  \Delta U^{(3)}_\alpha &=-4\pi G \mu v^{(0)}_\alpha,\label{eq:Uvec}\\
  v^{(2)\alpha}_{\phantom{(2)\alpha},\alpha} &=-
  \left({v^{(0)}}^2+\frac{p^{(0)}}{\mu}+
  2U\right)_{,\alpha}v^{(0)\alpha},\label{eq:continuity_equation}\\
\begin{split}\label{eq:pressure_gradient}
  \frac{p^{(2)}_{,\alpha}}{\mu} &= \left(\delta U+2\Phi +2 {v^{(0)}}^2 U
  +\frac{{p^{(0)}}^2}{2\mu^2}\right)_{,\alpha}-\\
  &2U^{(0)}{{v^{(0)}}^2}_{,\alpha}+4v^{(0)\beta}\left(U_{\alpha,\beta}-
  U_{\beta,\alpha}\right)+\\
  &v^{(0)\beta}\left(\left({v^{(0)}}^2+4U\right)
  v^{(0)}_{\phantom{(2)}\alpha,\beta}+\right.\\
  &\left.4v^{(0)}_{\alpha}U_{,\beta}\right)-v^{(0)\beta}
  v^{(2)}_{\phantom{(2)}\alpha,\beta}-v^{(2)\beta}
  v^{(0)}_{\phantom{(2)}\alpha,\beta},
\end{split}
\end{align}
\end{subequations}
where ${v^{(0)}}^2$ is the square of the Newtonian velocity field. Since the
spatial part of the metric is easily obtained and already incorporated in
Equation \eqref{eq:expansion of the metric} the sole equations that remain to be
solved in order to obtain the metric to 1 PN are the first two and the one which
determines $\delta U$. The other equations determine the 1 PN corrections to the
pressure, velocity field and the surface.

\subsection{The velocity field and the surface}

For the 1 PN correction of the surface we choose the
following ansatz
\begin{subequations}\label{eq:ansatz}
\begin{align}\label{eq:ansatz_surface}
\begin{split}
  S^{(2)}&=2\pi G \mu\left(  a_1^2\sum\limits_{i=1}^{2}
  S_i\left(\left(\frac{x^i}{a_i}\right)^2-
  \left(\frac{x^3}{a_3}\right)^2\right)+\right.\\
  &\left.S_3\left(x^1\right)^2\left(
  \frac{1}{3}\left(\frac{x^1}{a_1}\right)^2-
  \left(\frac{x^2}{a_2}\right)^2\right)+\right.\\
  &\left.S_4\left(x^2\right)^2\left(
  \frac{1}{3}\left(\frac{x^2}{a_2}\right)^2-
  \left(\frac{x^3}{a_3}\right)^2\right)+\right.\\
  &\left.S_5\left(x^3\right)^2\left(
  \frac{1}{3}\left(\frac{x^3}{a_3}\right)^2-
  \left(\frac{x^1}{a_1}\right)^2\right)\right)
\end{split}
\end{align}
and for the velocity
\begin{align}\label{eq:ansatz_velocity}
\begin{split}
  \frac{v^{(2)}_1}{\left(\pi G\mu\right)^{\tfrac 3 2}} &= x^2\left( a_1^2 w_1 +
  \hat q_1 \left(x^1\right)^2 +\right.\\
  &\quad\quad\quad\left. r_1 \left(x^2\right)^2 + t_1
  \left(x^3\right)^2\right),\\
  \frac{v^{(2)}_2 }{\left(\pi G\mu\right)^{\tfrac 3 2}}&= x^1 \left( a_2^2 w_2 
  + \hat q_2  \left(x^2\right)^2 +\right.\\
  &\quad\quad\quad\left. r_2 \left(x^1\right)^2 + t_2
  \left(x^3\right)^2\right),\\
  \frac{v^{(2)}_3 }{\left(\pi G\mu\right)^{\tfrac 3 2}}&= q_3 x^1 x^2 x^3.
\end{split}
\end{align}
\end{subequations}
In Paper II, both, $w_1$ and $w_2$, were not considered. In
\cite{Gurlebeck_2010}, we showed that these linear contributions can be used to
allow a rigidly rotating axisymmetric limit of the 1 PN Dedekind ellipsoids
coinciding with the 1 PN Maclaurin ellipsoids. Moreover, it seems natural to
allow a 1 PN contribution to the Newtonian constant $\Omega$ analogous to the
angular velocity in the case of rigid rotation, see Papers I. Furthermore, we
introduced the constant $\hat q_1=q_1+q$ and $\hat q_2=q_2-q$ compared to Paper
II. The ansatz for the surface is the same as in Paper I, i.e., it originates
from a Lagrangian displacement of all fluid elements such that the coordinate
volume remains constant. The Lagrangian displacement reads, cf. Equation (41) in
Paper II,
\begin{align}\label{eq:lagrangian_displacement}
\begin{split}
  \xi_\mu=&\frac{\pi G\mu a_1^2}{c^2} \sum\limits_{A=1}^{5} S_A \xi^{A}_\mu,\\
  \left(\xi_\mu^1\right)&=\left(x^1,0,-x^3\right),\quad 
  \left(\xi_\mu^2\right)=\left(0,x^2,-x^3\right),\\
  \left(\xi_\mu^3\right)&=\frac{1}{3a_1^2}
  \left(\left(x^1\right)^3,-3\left(x^1\right)^2x^2,0\right),\\
  \left(\xi_\mu^4\right)&=\frac{1}{3a_1^2}
  \left(0,\left(x^2\right)^3,-3\left(x^2\right)^2x^3\right),\\
  \left(\xi_\mu^5\right)&=\frac{1}{3a_1^2}
  \left(-3\left(x^3\right)^2x^1,0,\left(x^3\right)^3\right).
\end{split}
\end{align}

As was pointed out in \cite{Bardeen_1971}, it is more physical to fix parameters
with an immediate physical interpretation like the mass instead the coordinate
volume. However, by following Paper II, it is easier to compare our results with
theirs. As it was discussed in \cite{Chandrasekhar_1971a} and in Paper II, an
arbitrary contribution to the coordinate volume can be achieved by an additional
displacement of the form $\xi_6=\pi G\mu a_1^2c^{-2}S_6(x^1,x^2,x^3)$, whose
divergence does not vanish. Alternatively, one can arrive at the same result by
considering a different underlying Newtonian configuration, i.e. by substituting
in our final results $a_1$ by $a_1+2\pi G\mu a_1^2c^{-2}S_6$ while keeping
$\ba_2$ and $\ba_3$ fixed.

Note that higher order polynomials could also be allowed in the ansatz
\eqref{eq:ansatz}, which would eventually lead to a homogeneous system for their
coefficients. The form of the solution we use here is the minimal one needed to
satisfy the inhomogeneous equations and get a configuration that is
Dedekind-like in the sense discussed in Section \ref{sec:introduction}.

\subsection{The solution for the metric functions} 
\label{sec:metric_functions}

In this section, we will repeat the interior solution for the metric components
to an extent that is necessary to construct the exterior solution as well. Since
the changes in the ansatz \eqref{eq:ansatz} do not affect Equations \eqref{eq:1
PNPhi} and \eqref{eq:Uvec} their solutions are the same as obtained in Paper II.
The equations are of the type treated in \cite{Ferrers:1877}, namely they have a
polynomial density due to the form of the Newtonian solution
\eqref{eq:Newtonian_solution}. Thus, they can be expressed in terms of the
higher moments of the homogeneous density $D_{i_1i_2\ldots}$, cf.\ Appendix
\ref{sec:higher_moments}. With these, the solutions to Equations \eqref{eq:1
PNPhi} and \eqref{eq:Uvec} read
\begin{align}\label{eq:metric_interior}
\begin{split}
  \Phi&=\pi  G \mu  \left[U \left(\frac{3}{2} a_3^2 A_3+A_{\emptyset
  }\right)-\frac{5}{2} A_3 D_{33}\right.\\
  &\left.-\left(A_1+\frac{3}{2} \ba_3^2 A_3- \ba_2^2 \bar{\Omega
  }^2\right)D_{11}\right.\\
  &\left.-\left(A_2+\frac{3\ba_3^2}{2\ba_2^2}  A_3- \frac{1}{\ba_2^{2}
  }\bar{\Omega }^2\right)D_{22} \right],\\
  U_1&=-\frac{a_1}{a_2} \Omega  D_2,\quad U_2=\frac{a_2}{a_1} \Omega  D_1,\\
  U_3&=0.
\end{split}
\end{align} 

A similar approach can be taken for $\delta U$. It can be written as
\begin{align}\label{eq:deltaU_Poisson}
  \frac{\delta U}{c^2}=-G\frac{\partial}{\partial x^\mu} \int \frac{\mu
  \xi^\mu}{|\vec x-\vec x'|} d^3x',
\end{align}
cf.\ Equation (58) in \cite{Chandrasekhar_1967a}, i.e., as the sum of
derivatives of  Poisson integrals. As such $\delta U$ is not continuously differentiable
across the surface, which is a problem inherent to this coordinate system. Using
surface adapted Cartesian coordinates $y^\mu=x^\mu+\xi^\mu$ the metric
components are continuously differentiable. But we do not use them here but
rather follow Paper II.

The densities in the Poisson integrals \eqref{eq:deltaU_Poisson} are polynomial
for the Lagrangian displacement $\xi^\alpha$ \eqref{eq:lagrangian_displacement}.
Thus, the algorithm described in Section \ref{sec:exterior_solution} is
applicable. Taking the derivatives afterwards yields
\begin{align}\label{eq:Newtonian_potential_Pertubations}
\begin{split}
  \delta &U=-\mu^2 G^2 \pi \big(S_1a_1^2\left(D_{1,1}-D_{3,3}\right)+\\
  &S_2a_1^2\left(D_{2,2}-D_{3,3}\right)+\frac{S_3}{3}\left(D_{111,1}-3D_{112,2}\right)+\\
  &\frac{S_4}{3}\left(D_{222,2}-3D_{223,3}\right)+
  \frac{S_5}{3}\left(D_{333,3}-3D_{331,1}\right)\big).
\end{split}
\end{align}
Note that only six of the ten third order moments $D_{ijk}$ and only the
diagonal terms of the second order moments $D_{ij}$ are
necessary, cf.\ Equation \eqref{eq:metric_interior}. In order to fix the
constants $S_i$, we solve the Bianchi identity in Equation \eqref{eq:1
PN_equations} in the next section.

\subsection{Corrections to the pressure, the velocity field, and the
surface}\label{sec:matter_functions}

Although the changes in the ansatz will not change the calculations
fundamentally, we will describe it in more detail and give also intermediate
solutions and analytic expressions. The main reason is that we were unable to
reproduce the numerical data given in Paper II for the case $w_1=w_2=0$ (see the
discussion in \cite{Gurlebeck_2010} for further details). Thus, it might prove
helpful for rectifying this discrepancy or at the very least make our
calculations repeatable. The formulae, which are too lengthy for the text here,
can be found in Appendix \ref{appendix:explicit_results}.

Inserting our ansatz in the integrability condition for the gradient of the
pressure, i.e.\ Equation \eqref{eq:pressure_gradient}, and using the polynomial
structure for a comparison of coefficients yields the following solution:
\begin{align}\label{eq:solution_integrability_condition}
\begin{split}
  t_2&=\ba_2^2 t_1,\\
  r_2&=4  \left(\bar{a}_2^3-\bar{a}_2\right) \bar{\Omega }
  \left(\bar{B}_{112}+\bar{B}_{122}\right)+\left(\frac{1}{\bar{a}_2}
  -\bar{a}_2^3\right)  \bar{\Omega }^3\\
  &+\frac{1}{3} \left(\hat q_1\left(\bar{a}_2^2+2\right)+ \hat q_2 \left(2
  \bar{a}_2^2+1\right)-3 r_1 \bar{a}_2^2\right),\\
  q_3&=-4 \left(\bar{a}_2-\frac{1}{\bar{a}_2}\right) \bar{\Omega }
  \left(\bar{B}_{123}+\bar{B}_{13}\right).
\end{split}
\end{align}
Thus, the constant $q_3$ is already determined completely and independently of
the parameters $w_i$.

Repeating the same for the continuity Equation \eqref{eq:continuity_equation}
and using the results \eqref{eq:solution_integrability_condition} gives further
constraints on the constants
\begin{align}\label{eq:solution continuity equation}
\begin{split}
\hat q_2&=2 \left(\bar{a}_2-\frac{1}{\bar{a}_2}\right) \bar{\Omega}
\left(\bar{a}_2^2 \bar{B}_{123}+\bar{B}_{23}\right)\\
&+2\left(\bar{a}_2-\frac{1}{\bar{a}_2}\right) \bar{\Omega }^3-\hat q_1.
\end{split}
\end{align}

A further simplification is achieved by requiring the necessary condition that
the normal component of the velocity vanishes at the surface up to 1 PN. This
gives
%% The folowing two align environments belong together and were only split for
% type setting reasons
\begin{align*}
  r_1&=\frac{1}{\bar{a}_2^3}\left[\bar{\Omega } \bigg(\frac{1}{3
  \left(\bar{a}_2^2-1\right)}\left[3 (5\ba_2^2+1)(S_2-S_1)\right.\right.\\
  &\left.\left.-3(3\bar{a}_2^2-1)S_3+2\bar{a}_2^2(2\bar{a}_2^2+1)
  S_4\right]\right.\\
  &\left.-6\bar{a}_2^2\left(\bar{B}_{112}+\bar{B}_{122}\right)\right.\\
  &\left.- 2\left(2 \bar{a}_2^2+1\right)\left(\bar{a}_2^2\bar{B}_{123}+
  \bar{B}_{23}\right)\bigg)\right.\\
  &\left.-\frac{\left(w_1+w_2\right) \bar{a}_2 \left(5 \bar{a}_2^2+1\right)}{2 
  \left(\bar{a}_2^2-1\right)}-\frac{\left(5 \bar{a}_2^2+1\right) \bar{\Omega
  }^3}{2 }\right],
\end{align*}
\begin{align}\label{eq:normal_component_velocity}
\begin{split}
  t_1&=\frac{1}{2\bar{a}_2 \bar{a}_3^2}\left[ 2\bar{\Omega }
  \left(S_2-S_1-\bar{a}_2^2 S_4+\bar{a}_3^2 S_5\right.\right.\\
  &\left.\left.+2\left(\bar{a}_2^2-1\right)\left(
  \bar{B}_{123}+\bar{B}_{13}\right)\right)-\bar{a}_2\left(w_1+
  w_2\right)\right],\\
  \hat q_1&=\bar{\Omega } \Big(\frac{1}{
  \bar{a}_2\left(\bar{a}_2^2-1\right)}\left[3 (\ba_2^2+1)(S_2-S_1)\right.\\
  &\left.-(5\bar{a}_2^2-3)S_3+2\bar{a}_2^2S_4\right]-
  6\bar{a}_2\left(\bar{B}_{112}+\bar{B}_{122}\right)\\
  &- 2\frac{ \left( \bar{a}_2^2+2\right)}{\ba_2}\left(\bar{a}_2^2\bar{B}_{123}+
  \bar{B}_{23}\right)\Big)\\
  &-\frac{3\left(w_1+w_2\right)  \left( \bar{a}_2^2+1\right)}{2 
  \left(\bar{a}_2^2-1\right)}-\frac{\left( \bar{a}_2^2+5\right) \bar{\Omega
  }^3}{2 \bar{a}_2}.
\end{split}
\end{align}
Up to now, all coefficients entering the 1 PN corrections to the velocity field
can be given in terms of $w_i$ and $S_i$. To determine the surface coefficients
$S_i$, we have to impose that the pressure vanishes at the surface up to first
PN order, which leads to a linear system of equations:
\begin{align}\label{eq:surface_condition}
\sum\limits_{j=0}^{5}M_{ij}S_j=b_i^{(0)}+b_i^{(1)}w_1+b_i^{(2)}w_2=b_i.
\end{align}
We give the analytic and lengthy expressions of the coefficient matrix
$(M_{ij})$ and the inhomogeneity $(b_i)$ in Appendix
\ref{appendix:surface_condition}. The equations for $i=1,\ldots,5$ ensure that
the pressure at the surface is constant. Having solved those, the PN
contribution to the central pressure $p^{(2)}_C=S_0 a_1^4 \mu (\mu G)^2$ is
obtained using the equation for $i=0$ such that the pressure vanishes at the
surface.The parameters $S_i,~t_i,~r_i$ and $q_i$ are  plotted along the 1 PN 
Dedekind sequence in Figure \ref{Fig:parameters} in Appendix 
\ref{appendix:solution_plotted} using a parameterization described  in Section
\ref{sec:singularities}.

The singularity which was discovered in Paper II has its origin in a vanishing
determinant of the coefficient matrix $(M_{ij})$. How this singularity can be
removed is discussed in Section \ref{sec:singularities}. A solution $S_i$ of
Equation \eqref{eq:surface_condition} depends on the $w_i$. Hence, all constants
entering our ansatz \eqref{eq:ansatz} but $q_3$ are obtained in terms of the
$w_i$. The only requirement for the choice of the $w_i$ is that we have a
Dedekind-like configuration in the sense of Section \ref{sec:introduction} and
that the resulting surface \eqref{eq:ansatz} is still closed. The latter is just
a reformulation of the fact that the 1 PN corrections must be small compared to
the Newtonian quantities though offering here an explicit and necessary
criterion.

\subsection{Properties of the solution}\label{sec:properties}

In this section, we discuss some properties of the family of solutions described
in Sections \ref{sec:metric_functions} and \ref{sec:matter_functions}. The $w_i$
can be chosen independently for each $\ba_2$ along the family so that two free
functions $w_i(\ba_2)$ remain. We assume here continuous functions $w_i$ in
order to ensure a continuous family of 1 PN Dedekind ellipsoids. In
Section \ref{sec:singularities}, we determine the conditions for the $w_i$
imposed by the requirement that the 1 PN Dedekind ellipsoids have similar
properties as in the Newtonian case. We treated one of these requirements in
\cite{Gurlebeck_2010}, which ensures that the 1 PN Dedekind ellipsoids are
axially symmetric and rigidly rotating in the limit $\ba_2\to 1$ coinciding with
the 1 PN Maclaurin ellipsoids. This is achieved if $w_1(1)=-w_2(1)$ holds in the
limit. Analytic expressions can be found in that paper. Other restrictions do
not ensue from this property.

\subsubsection{The mass and the angular
momentum}\label{sec:mass_and_angular_momentum}

\begin{figure}[tbh!]
\raggedleft
\subfigure[\hspace{-1.2cm}]{\includegraphics[scale=0.365]{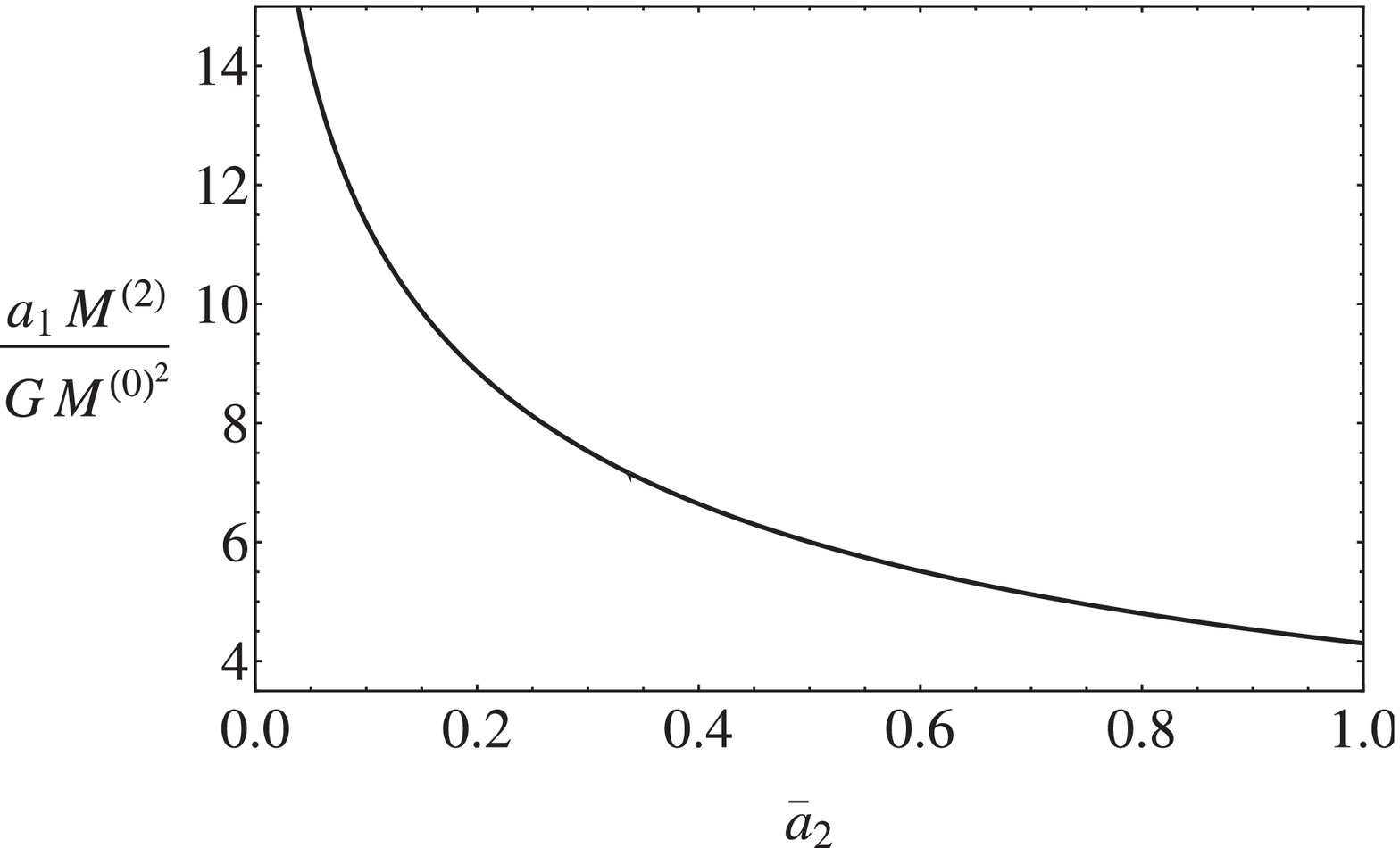}
\label{fig:Mass}
}\\ 
\subfigure[\hspace{-1.5cm}]{\includegraphics[scale=0.365]
{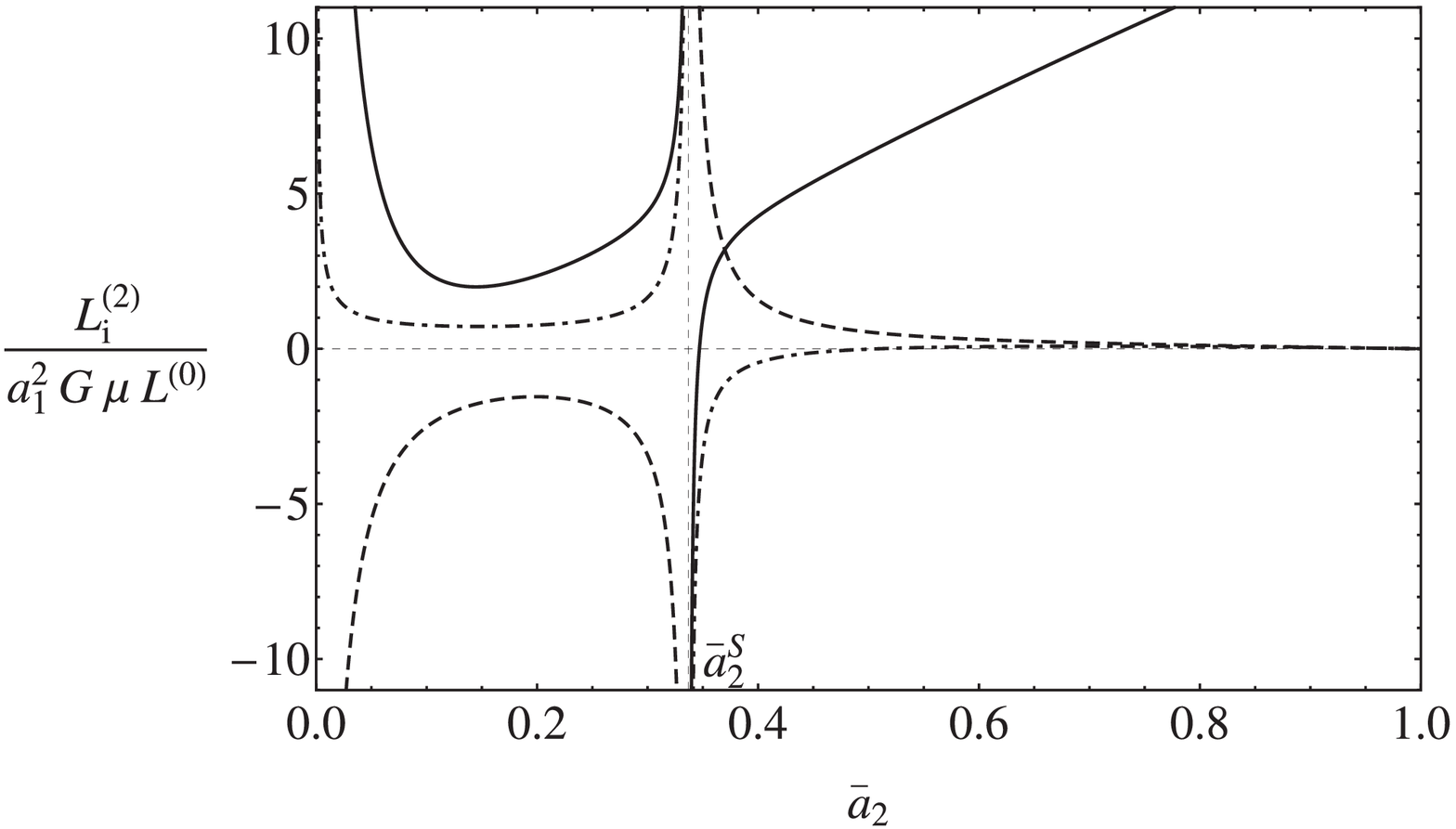}
\label{fig:AngularMomentum}
}
\caption{1 PN corrections to (a) the mass and (b) the angular
momentum.\label{fig:mass_angular_momentum}}
\end{figure}

Let us first characterize the 1 PN corrections by two physical parameters of the
Dedekind ellipsoids -- the mass and the angular momentum. According to the
definition in \cite{Chandrasekhar_1969a}, the 1 PN perturbation of the conserved
mass $M^{(2)}$ is
\begin{align}
  M^{(2)}=M^{(0)} \frac{\pi}{5}G\mu a_1^2 (12 \bA_{\emptyset} + (1 + \ba_2^2)
\bB_{12}).
\end{align}
This is independent of the choice of $w_i$. The angular momentum of the
Newtonian solution as well as the 1 PN Dedekind ellipsoids points in the
$x^3$-direction. The first evaluates to
\begin{align}\label{eq:angular_momentum_Newtonian}
   L^{(0)}=\frac{8\pi}{15}\mu\Omega a_1^2a_2^2a_3,
\end{align}
whereas the latter is rather lengthy and is shown in Appendix
\ref{appendix:angular_momentum}. Here, we only present
the plot, cf. Figure \ref{fig:mass_angular_momentum}. $L^{(2)}$ is linear in
the $w_i$ as are all other quantities, which we investigate. Thus, we depict for all
constants, say $L^{(2)}$, the coefficients in front of the $w_i$, i.e.,
$L^{(2)}=L^{(2)}_{0}+ L^{(2)}_{1}w_1+ L^{(2)}_{2}w_2$. We use a solid line for
$L^{(2)}_0$, a dashed line for $L^{(2)}_1$ and a dot-dashed line for
$L^{(2)}_2$. The part $L^{(2)}_0$ equals the respective constants in Paper II,
though as discussed in length in \cite{Gurlebeck_2010} the numerical
values do not agree.

\subsubsection{The singularity}\label{sec:singularities}

The singularity at\footnote{Subsequently, we truncate numerical values to six
digits.} $\ba_2^s=0.33700003168\ldots$, cf.\ Figure \ref{fig:AngularMomentum}, is the
one already discovered in Paper II. The importance of singularities in 1 PN
approximations of equilibrium figures and the issues with the 1 PN Dedekind
sequence of that paper was discussed in Section \ref{sec:introduction}. Since we
are able to introduce singularities in our 1 PN Dedekind solutions at arbitrary
points $\ba_2$ via $w_i\left(\ba_2\right)$, it is obvious that not all
singularities are necessarily at physically distinguished points. That we can
use the $w_i\left(\ba_2\right)$ to remove the singularity at $\ba_2^S$ for
\emph{all} physical quantities is shown in this section.

Evaluating the determinant of the coefficient matrix $(M_{ij})$, cf. Equation
\eqref{eq:surface_condition} and Appendix \ref{appendix:surface_condition},
numerically shows that it changes sign at $\ba_2^s$. One of the eigenvalues, say
$\lambda$, of the coefficient matrix $(M_{ji})$ vanishes there. Multiplying
Equation \eqref{eq:surface_condition} with the eigenvector $(\beta_i)$ to the
eigenvalue $\lambda$ of the transposed matrix yields the condition
\begin{align}\label{eq:sing_cond}
\begin{split}
&\sum\limits_{i=1}^3\beta_ib_i=\\
&0.083600 - 0.235534 w_1 + 0.099994 w_2=0.
\end{split}
\end{align}
In Paper II, whose results are obtained by setting $w_1=w_2=0$, Equation
\eqref{eq:sing_cond} could not be satisfied. Hence, a singularity is inevitable
there in lieu of our approach. We have a more general
inhomogeneity in Equation \eqref{eq:surface_condition}, cf. the $b^{1/2}_i$.
Choosing at $\ba_{2}^s$
\begin{align}\label{eq:sing_value_w1}
w^s_1 = 0.354937 + 0.424544 w_2,
\end{align}
Equation \eqref{eq:sing_cond} is identically satisfied and the remaining system
of four equations in Equations \eqref{eq:surface_condition} is regular. To ensure that
this holds in the limit $\ba_2\to\ba_2^s$ as well, higher orders have to be
taken into account. However, as one can readily check numerically, the equation
$\sum_{i,j}\frac{\beta_iM_{ij}S_j}{\lambda}=\sum_{i}\tfrac{\beta_ib_i}{\lambda}$
is well-defined in the limit $\ba_2\to\ba_2^s$ if $w_1-w_1^s\in O(\lambda)$. Thus, this equation
can be used instead of, for instance, $\sum_jM_{1j}S_j=b_1$. The resulting
coefficient matrix is now regular and the solution finite. Let us introduce instead of $w_1$
a new parameter defined by $w_1=w_1^s(w_2)+\lambda \hat w_1$, where $\hat w_1$
and $w_2$ are arbitrary. Then, the surface coefficients at $\ba_{2}^s$ evaluate
to
\begin{align}
\begin{split}
  S^s_1&=0.136453 - 0.243073 w_2 + 0.054186 \hat w_1,\\
  S^s_2&=- 0.195876 + 0.154884 w_2 - 0.044769 \hat w_1,\\
  S^s_3&= -0.119902 + 0.212999 w_2 + 0.221751 \hat w_1,\\
  S^s_4&= -1.393285 + 0.412379 w_2 + 0.429324 \hat w_1,\\
  S^s_5&= 4.761466 - 1.470110 w_2 - 1.530518 \hat w_1.
\end{split}
\end{align}
Since these parameters enter the velocity field linearly and their coefficients
in Equations \eqref{eq:solution_integrability_condition}--\eqref{eq:normal_component_velocity}
are well-defined at $\ba_2^s$, the entire solution is regular. Only Equation
\eqref{eq:sing_value_w1} is obtained as an extra condition for $\ba_2\to
\ba_2^s$. As an example, the angular momentum reads under this constraint
\begin{align}
\begin{split}
  \frac{L_s^{(2)}}{a_1^7 G^{\frac{3}{2}} \mu^{\frac{5}{2}}}=0.177158 - 0.005261
  \hat w_1 + 0.009930 w_2.
\end{split}
\end{align}
In contrast to Figure \ref{fig:AngularMomentum}, this is finite. Equation
\eqref{eq:sing_value_w1} is not a good parametrization if $\ba_2\to 0$ or
$\ba_2\to 1$. For the axially symmetric limit, a good parametrization close to
$\ba_2=1$ was discussed in \cite{Gurlebeck_2010} and the limit $\ba_2\to 0$ is
treated in detail in Appendix \ref{sec:limit_rod}. These two limits yield
additional constraints for $w_i(\ba_2)$ in the neighborhood of the respective
points that can be satisfied simultaneously.

\subsubsection{The surface and the gravitomagnetic effect}\label{sec:gravitomagnetic}

Which shapes can we expect for the 1 PN Dedekind ellipsoids if no additional
constraints are given? The gravitomagnetic effect, i.e., parallel matter
streams repel each other, is already included in a 1 PN approximation. Thus, one
should find that the 1 PN Dedekind ellipsoid is elongated\footnote{Note that the
qualitative picture does not change if one evaluates proper distances instead of
coordinate distances as is done here for simplicity.} in the $x^3$-direction
compared to the Newtonian figure, since all the Newtonian streams are along
ellipses in the same direction in parallel planes. This qualitative argument is
corroborated by the Maclaurin ellipsoids, which are also elongated in this
direction close to the bifurcation point, see \cite{Petroff_2003}. Moreover, the
matter streams in the $x^2$-direction for $x^1>0$ are all parallel and
anti-parallel to those in $x^1<0$. The latter are on average at a larger
distance. Thus, the repulsive effect should prevail and in this direction too we
have an elongation. In the $x^1$-direction, a similar argument holds. Which
effect is stronger, the repulsion in the $x^2$-direction or in the
$x^1$-direction, depends on the choice of $w_i$.  However, this conclusion
disregards the pressure entirely and can only provide a rough idea for the
resulting 1 PN shape. More importantly, the Lagrangian displacement
\eqref{eq:lagrangian_displacement} yields a vanishing 1 PN contribution to the
coordinate volume and, thus, it does not allow for an elongation of the
ellipsoid in all directions. In fact, we always observe elongations in the
$x^3$-direction, and deformation in the other directions can be adjusted with
different choices of the $w_i$. The three qualitatively different case are
depicted in Figure \ref{fig:surface} in Appendix \ref{appendix:solution_plotted}.

\subsubsection{The motion of the fluid}

To explicate the inner motion of the 1 PN Dedekind ellipsoids, we integrated the
velocity field of the fluid \eqref{eq:ansatz} numerically and discuss here the
trajectories of a generic fluid element. In the Newtonian Dedekind ellipsoid,
all fluid elements move along ellipses that are in planes with constant $x^3$.
The time of revolution coincides for all fluid elements. In the 1 PN
approximation to these figures of equilibrium, the trajectories are distorted
ellipses that are still closed and the motion is not any longer confined to
planes with constant $x^3$. The latter follows directly from our ansatz in
Equation \eqref{eq:ansatz_velocity}. The velocity in the $x^3$-direction
vanishes at the coordinate planes so that no fluid element moves from the upper
half of the 1 PN ellipsoid ($x^3>0$) to the lower ($x^3<0$) and vice versa. It
also changes sign when the particle crosses the other coordinate planes.
Moreover, the motion in the $x^3$-direction is periodic with half the time of
revolution as period. Furthermore, the time of revolution of the fluid elements
depends now on the starting point. The trajectories for particles at the
surface, which stay there during their motion, are exemplary and can be inferred
partially from Figure \ref{fig:surface}.

\begin{acknowledgements}
We gratefully acknowledge discussions with M. Ansorg, J. Bi\v c\'ak, J.
Friedman and R. Meinel. NG was financially supported by the Grants No. GAUK
22708 and GA\v CR 205/09/H033 and DP by the Deutsche Forschungs\-gemein\-schaft
as part of the project ``Gravitational Wave Astronomy'' (SFB/TR7B1). Moreover,
NG gratefully acknowledges support from the DFG within the Research Training
Group 1620 ``Models of Gravity''.
\end{acknowledgements}

%\bibliographystyle{apj}
%\bibliography{References}

\appendix

\section{Higher moments}\label{sec:higher_moments}

The interior solutions for the higher moments of the homogeneous mass
distribution were obtained in \cite{Ferrers:1877}. They are given in
\cite{Chandrasekhar_1987} and we repeat them here:
\begin{align*}
\begin{split}
  D_i&=\pi G \mu a_i^2 x^i\left(A_i-\sum\limits_{l=1}^{3}A_{il}
  \left(x^l\right)^2\right),\\
  D_{ij}&=\pi G \mu a_i^2\left(a_j^2 \left(A_{ij}-\sum\limits_{l=1}^{3}A_{ijl}
  \left(x^l\right)^2\right) x^i x^j +\frac 1 4
  \delta_{ij}\left(B_i-2\sum\limits_{l=1}^{3}B_{il}
  \left(x^l\right)^2+\sum\limits_{l=1}^{3}\sum\limits_{k=1}^{3}B_{ilk}
  \left(x^l\right)^2\left(x^k\right)^2\right)\right),\\
  D_{ijk}&=\pi G \mu a_i^2a_j^2a_k^2\left(A_{ijk}-\sum\limits_{l=1}^3A_{ijkl}
  \left(x^l\right)^2\right)x^ix^jx^k+\frac{1}{4}\left(V_{ijk}+V_{jki}+
  V_{kij}\right),\\
  V_{ijk}&=a_i^2a_j^2\delta_{jk}\left(B_{ij}-\sum\limits_{l=1}^3\left(2B_{ijl}-
  \sum\limits_{m=1}^3B_{ijlm}\left(x^m\right)^2\right)
  \left(x^l\right)^2\right)x^i.
\end{split}
\end{align*}

\section{Explicit analytical results}\label{appendix:explicit_results}

We present the results that are too lengthy for the main text.

\subsection{The angular momentum}\label{appendix:angular_momentum}

In Section \ref{sec:mass_and_angular_momentum}, we discussed the 1 PN
contribution to the angular momentum. We gave the analytic expression for the
Newtonian angular momentum in Equation \eqref{eq:angular_momentum_Newtonian} and
the plot for the 1 PN contribution, cf.\ Figure \ref{fig:AngularMomentum}. Here, we
provide the analytic expression for $L^{(2)}$, too:
\begin{align}
\begin{split}
  L^{(2)}=&
  L^{(0)} \frac{\pi  G}{7} \mu \left(-26  a_1^2  A_1- 
  \frac{1}{\left(a_1^2-a_2^2\right)}\left(  \left(5 a_1^2+ 
  19 a_2^2\right) a_1^2  S_1+ \left(19 a_1^2 + 
  5 a_2^2\right) a_1^2 S_2+
  4  \left(a_1^2-3 a_2^2\right)\times\right.\right.\\
  &\left.\left. a_1^2 S_3+a_2^2 \left(7 a_1^2+a_2^2\right)
  S_4\right)-a_3^2 S_5 +2 \left(3 a_1^4 A_{11}+
  2 a_1^2 \left(a_2^2 \left(6  \left(B_{112}+
  B_{122}\right)+A_{12}+3 B_{123}\right)+
  \right.\right.\right.\\
  &\left.\left.\left.4 B_{12}+3 B_{23}\right)+
  a_2^2 \left(8 B_{12}+3 a_2^2  \left(2 B_{123}+
  A_{22}\right)+6 B_{23}\right)+
  a_3^2 \left(a_1^2 A_{13}+a_2^2 A_{23} -
  2 A_3\right)+\right.\right.\\
  &\left.\left.21 A_{\emptyset }\right)-26 a_2^2 
  A_2\right)-   \frac{4 \pi ^{5/2}a_3 a_1^3 a_2^3 
  \mu  (G \mu )^{3/2}}{105 \left(a_1^2-
  a_2^2\right)}\left(
  \left(19 a_1^2+5 a_2^2 \right) w_1 - \left(5 
  a_1^2+19 a_2^2 \right) w_2\right).
\end{split}
\end{align}

\subsection{The surface condition}\label{appendix:surface_condition}

Our numerical results and the numerical results in Paper II do not coincide as
discussed in detail in \cite{Gurlebeck_2010}. We believe that there is a problem
in the numerical evaluation of the right hand side of Equation
\eqref{eq:surface_condition} in Paper II. However, this could not be explicitly
seen since those expressions were not given. We provide the lengthy analytical
expression for Equation \eqref{eq:surface_condition}, which we obtained and used
in all numerical considerations in our text. To shorten the results, we
introduce a third kind of index symbols
\begin{align}
\begin{split}
  C_{i_1\ldots i_k}=\int\limits_{0}^{\infty}\frac{u^2}{H(u)z_{i_1} z_{i_2}\cdots
  z_{i_k}} d u,\quad 
  C_{i_1 \ldots i_k}(a_1,a_2,a_3)&=a_1^{6-2k} C_{i_1 \ldots i_k}(1,\ba_2,\ba_3)=a_1^{6-2k}\bar{C}_{i_1\ldots i_k}
\end{split}
\end{align}
with the same meaning of $H(u)$ and $z_i$ as in Equation \eqref{eq:abbreviation index
symbols}. The coefficient matrix $\left(M_{ij}\right)$ reads
\begin{flalign}\label{eq:coeff_mat}
\begin{split}
  M_{00}&=\frac{1}{\pi ^2}\\
  M_{01}&=3 \bar{a}_3^2 \bar{B}_{33}-\bar{B}_{13}\\
  M_{02}&=3 \bar{a}_3^2 \bar{B}_{33}-\bar{a}_2^2 \bar{B}_{23}\\
  M_{03}&=\frac{1}{4} \left(\bar{a}_2^2-1\right) \left(\bar{C}_{112}-2
\bar{a}_3^2 \bar{C}_{1123}+\bar{a}_3^4 \bar{C}_{11233}\right)
\end{split}
\end{flalign}
\begin{align*}
  M_{04}=&-\frac{\bar{a}_2^2}{4}  \left(\left(\bar{a}_2^2-\bar{a}_3^2\right)
  \left(\bar{C}_{223}-2 \bar{a}_3^2 \bar{C}_{2233}+\bar{a}_3^4
  \bar{C}_{22333}\right)+4 \bar{a}_3^4 \bar{C}_{2333}\right)\\
  M_{05}=&-\frac{\bar{a}_3^2 }{12} \left(\bar{a}_3^2 \left(11
  \bar{C}_{133}+\bar{a}_3^2 \left(-10 \bar{C}_{1333}+11
  \left(\bar{a}_3^2-1\right) \bar{C}_{13333}+24 \bar{a}_3^2
  \bar{B}_{3333}\right)+14 \bar{C}_{1333}\right)-3 \bar{C}_{133}\right)\\
  M_{10}=&M_{20}=M_{30}=M_{40}=M_{50}=0\\
  M_{11}=&\frac{2}{\bar{a}_2^2} \left(\left(\bar{a}_3^2-1\right)
  \left(\bar{a}_3^2-\bar{a}_2^2\right) \bar{C}_{1233}-2 \bar{a}_3^2
  \bar{B}_{33}\right)\\
  M_{12}=&\frac{2}{\bar{a}_2^2} \left(3 \left(\bar{a}_2^2-\bar{a}_3^2\right){}^2
  \bar{C}_{2233}-4 \bar{a}_3^2 \bar{B}_{23}\right)\\
  M_{13}=&\frac{1}{\bar{a}_2^2}\left(\left(\left(\bar{a}_2^2-1\right)
  \bar{a}_3^2+\bar{a}_2^2\right) \bar{C}_{1123}-\left(\bar{a}_2^2-1\right)
  \bar{a}_3^4 \bar{C}_{11233}-3 \bar{a}_2^4 \bar{C}_{1223}\right)\\
  M_{14}=&5 \bar{a}_2^4 \bar{C}_{2223}-4 \bar{a}_3^4 \bar{a}_2^4
  \bar{A}_{2233}-2 \left(\bar{a}_2^4+\bar{a}_3^2 \bar{a}_2^2+\bar{a}_3^4\right)
  \bar{C}_{2233}+5 \bar{a}_3^4 \bar{C}_{2333}+4 \bar{a}_3^2 \bar{A}_3\\
  M_{15}=&\frac{\bar{a}_3^2 }{3 \bar{a}_2^2}\left(3 \bar{a}_3^4 \left(-2
  \bar{C}_{1233}+3 \bar{C}_{1333}+5 \bar{C}_{2333}-9
  \bar{C}_{3333}\right)+\bar{a}_3^2 \left(3 \left(\bar{a}_2^2-1\right)
  \bar{C}_{1233}+6 \bar{C}_{1333}+8 \bar{A}_3\right)-\right.\\
  &\left.3 \bar{a}_2^2 \bar{C}_{1233}-8 \bar{a}_3^8 \bar{A}_{3333}\right)\\
  M_{21}=&-M_{22}=\frac{48 \left(2 \bar{a}_2^2+1\right)
  \bar{B}_{12}}{\bar{a}_2^2 \left(\bar{a}_2^2-1\right)}\\
  M_{23}=&\left(\left(\frac{12}{\bar{a}_2^2}-6\right) \bar{a}_3^2-6\right)
  \bar{C}_{11223}+\frac{6 \left(\bar{a}_2^2-1\right) \bar{a}_3^4
  \bar{C}_{11233}}{\bar{a}_2^4}+\frac{\left(48-96 \bar{a}_2^2\right)
  \bar{B}_{12}}{\bar{a}_2^2-\bar{a}_2^4}+30 \left(\bar{a}_2^2-\bar{a}_3^2\right)
  \bar{C}_{12223}\\
  M_{24}=&-\frac{16 \left(\bar{a}_2^2+2\right) \bar{B}_{12}}{\bar{a}_2^2-1}+16
  \bar{a}_2^6 \bar{A}_{2222}+54 \left(\bar{a}_3^2-\bar{a}_2^2\right) \bar{a}_2^2
  \bar{C}_{22223}+12 \left(2 \bar{a}_2^4-\bar{a}_3^2
  \bar{a}_2^2-\bar{a}_3^4\right) \bar{C}_{22233}+\\
  &48 \bar{a}_3^4 \bar{a}_2^2 \bar{A}_{2233}+30 \bar{a}_3^4
  \left(\frac{\bar{a}_3^2}{\bar{a}_2^2}-1\right) \bar{C}_{22333}-\frac{64
  \bar{a}_3^2 \bar{A}_3}{\bar{a}_2^2}\\
  M_{25}=&\frac{2}{\bar{a}_2^4}\left(-3\bar{a}_2^4 \left(\bar{a}_3^2-1\right)
  \bar{a}_3^2 \bar{C}_{12233}+3 \left(4 \bar{a}_3^4+\bar{a}_3^2-\bar{a}_2^2
  \left(3 \bar{a}_3^2+2\right)\right) \bar{a}_3^4 \bar{C}_{12333}+\right.\\
  &\left.27 \left(\bar{a}_2^2-\bar{a}_3^2\right) \bar{a}_3^6 \bar{C}_{23333}-8
  \bar{a}_3^4 \bar{A}_3+8 \bar{a}_3^{10} \bar{A}_{3333}\right)\\
  M_{31}=&-2 \left(\left(\bar{a}_3^2-1\right) \left(3 \bar{B}_{113}+\bar{a}_3^2
  \bar{B}_{133}\right)+4 \bar{a}_3^2 \bar{B}_{33}\right)\\
  M_{32}=&2 \left(\left(\bar{a}_3^2-1\right)
  \left(\bar{a}_3^2-\bar{a}_2^2\right) \bar{C}_{1233}-2 \bar{a}_3^2
  \bar{B}_{33}\right)\\
  M_{33}=&-5 \left(\bar{a}_2^2-1\right) \bar{C}_{1112}+5
  \left(\bar{a}_2^2-1\right) \bar{a}_3^2
  \bar{C}_{11123}+\left(\left(\bar{a}_3^2+4\right)
  \bar{a}_2^2-\bar{a}_3^2-2\right) \bar{C}_{1123}-\left(\bar{a}_2^2-1\right)
  \bar{a}_3^4 \bar{C}_{11233}\\
  M_{34}=&\left(\bar{a}_2^4+\left(\bar{a}_2^2-1\right) \bar{a}_3^2
  \bar{a}_2^2\right) \bar{C}_{1223}-3 \bar{a}_2^2 \bar{a}_3^4
  \bar{C}_{1233}+\bar{a}_2^2 \left(\bar{a}_3^2-2 \bar{a}_2^2\right) \bar{a}_3^2
  \bar{C}_{2233}+5 \bar{a}_2^2 \bar{a}_3^4 \bar{C}_{2333}\\
  M_{35}=&-3 \bar{a}_3^2 \bar{B}_{113}+3 \left(-2
  \bar{a}_3^4+\bar{a}_3^2+1\right) \bar{a}_3^4 \bar{C}_{11333}+2
  \left(\bar{a}_3^2+1\right) \bar{a}_3^4 \bar{B}_{133}+\left(6
  \bar{a}_3^2-7\right) \bar{a}_3^6 \bar{B}_{1333}+\frac{20}{3} \bar{a}_3^4
  \bar{A}_3-\\
  &\frac{20}{3} \bar{a}_3^8 \bar{A}_{333}-5 \bar{a}_3^6
  \bar{B}_{333}+\frac{35}{3} \bar{a}_3^8 \bar{B}_{3333}\\
  M_{41}=&-M_{42}=\frac{48 \bar{B}_{12}}{\bar{a}_2^2-1}\\
  M_{43}=&10 \left(\bar{a}_2^2-1\right) \bar{C}_{11122}-\frac{10
  \left(\bar{a}_2^2-1\right) \bar{a}_3^2 \bar{C}_{11123}}{\bar{a}_2^2}-8
  \bar{a}_2^2 \bar{A}_{1122}-\frac{2 \left(\bar{a}_2^2-1\right)
  \left(\bar{a}_2^2 \left(3 \bar{a}_3^2+2\right)-2 \bar{a}_3^2\right)
  \bar{C}_{11223}}{\bar{a}_2^2}+\\
  &\frac{2 \left(\bar{a}_2^2-1\right) \bar{a}_3^4
  \bar{C}_{11233}}{\bar{a}_2^2}+\frac{16 \bar{B}_{12}}{\bar{a}_2^2-1}+\frac{8
  \bar{a}_3^2 \bar{A}_3}{\bar{a}_2^2}\\
  M_{44}=&-\frac{16 \bar{a}_2^2 \bar{B}_{12}}{\bar{a}_2^2-1}+\left(2 \left(3
  \bar{a}_3^2-5\right) \bar{a}_2^4+4 \bar{a}_3^2 \bar{a}_2^2\right)
  \bar{C}_{12223}-4 \left(\bar{a}_3^2-1\right) \bar{a}_2^4 \bar{C}_{12233}+6
  \bar{a}_3^4 \left(\bar{a}_3^2-1\right) \bar{C}_{12333}+\\
  &4 \bar{a}_3^2 \left(\bar{a}_2^2-\bar{a}_3^2\right) \bar{a}_2^2
  \bar{C}_{22233}+8 \bar{a}_3^4 \bar{a}_2^4 \bar{A}_{2233}+4 \bar{a}_3^4
  \left(\bar{a}_3^2-\bar{a}_2^2\right) \bar{C}_{22333}-8 \bar{a}_3^2 \bar{A}_3\\
  M_{45}=&6 \bar{a}_3^2 \bar{B}_{1123}-\frac{12 \left(\bar{a}_3^2-1\right)
  \bar{a}_3^4 \bar{B}_{1133}}{\bar{a}_2^2}+\frac{18 \left(\bar{a}_3^2-1\right)
  \bar{a}_3^4 \bar{C}_{11333}}{\bar{a}_2^2}-4
  \left(\bar{a}_3^6+\bar{a}_3^4\right) \bar{B}_{1233}+\frac{8 \bar{a}_3^8
  \bar{A}_{1333}}{\bar{a}_2^2}+\\
  &\frac{2 \left(4 \bar{a}_3^8+\bar{a}_3^6\right)
  \bar{B}_{1333}}{\bar{a}_2^2}+10 \bar{a}_3^6 \bar{B}_{2333}-\frac{40
  \bar{a}_3^4 \bar{A}_3}{3 \bar{a}_2^2}+\frac{16 \bar{a}_3^{10}
  \bar{A}_{3333}}{3 \bar{a}_2^2}-\frac{18 \bar{a}_3^8
  \bar{B}_{3333}}{\bar{a}_2^2}\\
  M_{51}=&-M_{52}=\frac{48 \left(\bar{a}_2^2+2\right)
  \bar{B}_{12}}{\bar{a}_2^2-1}\\
  M_{53}=&16 \bar{A}_{1111}+54 \left(\bar{a}_2^2-1\right) \bar{C}_{11112}-12
  \left(\left(\bar{a}_3^2+4\right) \bar{a}_2^2-3 \bar{a}_3^2-2\right)
  \bar{C}_{11123}+6 \left(\bar{a}_2^2-1\right) \bar{a}_3^4 \bar{C}_{11233}+\\
  &\frac{16 \left(5 \bar{a}_2^2-2\right) \bar{B}_{12}}{\bar{a}_2^2-1}-16
  \bar{a}_3^2 \bar{A}_3\\
  M_{54}=&\left(6 \bar{a}_2^2 \bar{a}_3^2-6 \bar{a}_2^4\right)
  \bar{C}_{11223}-\frac{48 \bar{a}_2^2 \bar{B}_{12}}{\bar{a}_2^2-1}+12
  \bar{a}_2^2 \left(\bar{a}_2^2-\bar{a}_3^2\right) \bar{a}_3^2
  \bar{C}_{12233}+24 \bar{a}_2^2 \left(\bar{a}_3^2-1\right) \bar{a}_3^4
  \bar{C}_{12333}+\\
  &\left(6 \bar{a}_2^2 \bar{a}_3^6-6 \bar{a}_2^4 \bar{a}_3^4\right)
  \bar{C}_{22333}\\
  M_{55}=&-30 \left(\bar{a}_3^2-1\right) \bar{a}_3^2 \bar{C}_{11133}+12 \left(6
  \bar{a}_3^4-5 \bar{a}_3^2-1\right) \bar{a}_3^4 \bar{C}_{11333}+48 \bar{a}_3^8
  \bar{A}_{1333}-48 \left(\bar{a}_3^2-1\right) \bar{a}_3^6 \bar{B}_{1333}-\\
  &54 \left(\bar{a}_3^2-1\right) \bar{a}_3^6 \bar{C}_{13333}-64 \bar{a}_3^4
  \bar{A}_3+16 \bar{a}_3^{10} \bar{A}_{3333}.
  \end{align*}
The inhomogeneities are given by
\begin{flalign}\label{eq:inhom}
\begin{split}
  b_0&=-\frac{1}{4} \left(-2 \bar{C}_{13} \left(\bar{A}_1-2 \bar{a}_2^2
  \bar{B}_{12}\right)-\bar{a}_3^2 \left(-2 \bar{A}_1
  \left(\bar{C}_{133}+\bar{C}_{233}\right)+\bar{A}_3 \left(3
  \left(\bar{B}_1+\bar{B}_2\right)+\bar{C}_{33}\right)+\right.\right.\\
  &\left.\left.2 \bar{B}_{12} \left(\bar{a}_2^2 \left(2 \bar{C}_{133}-\bar{C}_{233}\right)+3 \bar{C}_{233}\right)\right)-2 \bar{C}_{23} \left(\bar{A}_1+\left(\bar{a}_2^2-3\right) \bar{B}_{12}\right)+\bar{a}_3^6 \bar{A}_3 \left(3
  \left(\bar{B}_{133}+\bar{B}_{233}\right)-16 \bar{B}_{333}\right)+\right.\\
  &\left.\bar{a}_3^4 \bar{A}_3 \left(6
  \left(\bar{C}_{133}+\bar{C}_{233}\right)-8 \bar{A}_3+9 \bar{C}_{333}\right)+8 \bar{A}_{\emptyset }^2\right)\\
  b_1&=-\frac{1}{a_2^2}\left(\bar{A}_1 \left(2 \bar{a}_2^2 \bar{C}_{123}-2
  \bar{a}_3^2 \bar{C}_{133}\right)+\bar{B}_{12} \left(-8
  \left(\bar{a}_2^2-1\right) \bar{a}_3^2 \bar{B}_{123}-4 \bar{a}_2^4
  \bar{C}_{123}-4 \left(\bar{a}_2^2-2\right) \bar{a}_3^2 \bar{B}_{13}-4
  \bar{a}_3^4 \bar{a}_2^2 \bar{B}_{133}+\right.\right.\\
  &\left.\left.8 \bar{a}_2^4 \bar{B}_{223}-4 \bar{a}_2^2 \bar{C}_{223}+16
  \bar{a}_3^2 \bar{a}_2^2 \bar{A}_{23}+4 \bar{a}_3^2 \bar{C}_{233}+8
  \left(\bar{a}_2^2-\bar{a}_3^2\right) \bar{A}_3-16
  \bar{a}_2^2\right)+\bar{A}_{\emptyset } \left(8 \bar{a}_2^2 \bar{B}_{12}+4
  \left(\bar{a}_3^2-\bar{a}_2^2\right) \bar{B}_{23}\right)+\right.\\
  &\left.\bar{A}_3 \left(3 \bar{a}_2^2 \bar{a}_3^2 \bar{C}_{123}-3 \bar{a}_3^4
  \bar{C}_{133}-6 \bar{a}_2^4 \bar{a}_3^2 \bar{B}_{223}+3 \bar{a}_2^2
  \bar{a}_3^2 \bar{C}_{223}+6 \left(\bar{a}_3^2-\bar{a}_2^2\right) \bar{a}_3^2 
  \bar{B}_{23}+\left(5 \bar{a}_2^2 \bar{a}_3^2-3 \bar{a}_3^4\right)
  \bar{C}_{233}+\right.\right.\\
  &\left.\left.10  \left(\bar{a}_2^2-\bar{a}_3^2\right) \bar{a}_3^6
  \bar{B}_{2333}+10 \bar{a}_3^6 \bar{B}_{333}-5 \bar{a}_3^4
  \bar{C}_{333}\right)+\bar{A}_2 \left(-4 \bar{a}_2^6 \bar{B}_{223}+2
  \bar{a}_2^4 \bar{C}_{223}-2 \bar{a}_3^2 \bar{a}_2^2
  \bar{C}_{233}\right)+\left(w_2-w_1\right) \bar{a}_2^3 \bar{\Omega }\right)\\
  b_2&=-\frac{6}{a_2^4} \left(2 \bar{A}_1 \left(\bar{a}_2^2 \left(4
  \bar{B}_{12}+\bar{a}_3^2 
  \left(\bar{C}_{1223}+\bar{C}_{1233}\right)\right)-\bar{a}_2^4
  \bar{C}_{1223}-\bar{a}_3^4 \bar{C}_{1233}\right)+4 \bar{B}_{12}
  \left(\bar{a}_2^4 \left(6 \bar{B}_{112}+6
  \bar{B}_{122}-\right.\right.\right.\\
  &\left.\left.\left.2 \bar{a}_3^2  \left(\bar{B}_{123}+2 \bar{A}_{223}\right)+2
  \bar{B}_{123}+7 \bar{B}_{222}+4 \bar{B}_{23}\right)+\bar{a}_2^6
  \left(\bar{B}_{122}+4 \bar{B}_{123}+2 \bar{A}_{222}\right)+\bar{a}_2^2 \left(2
  \bar{a}_3^2 \left(2\left(\bar{B}_{123}+\bar{B}_{13}\right)-
  \right.\right.\right.\right.\\
  &\left.\left.\left.\left.5 \bar{B}_{223}\right)+\bar{a}_3^4  \bar{B}_{133}+2
  \bar{B}_{23}\right)+\bar{a}_3^2 \left(-4
  \left(\bar{B}_{123}+\bar{B}_{13}\right)+\bar{a}_3^2 \bar{B}_{233}+4
  \bar{A}_3\right)\right)+2 \bar{a}_2^2 \bar{A}_2
  \left(\left(\bar{a}_2^2-\bar{a}_3^2\right) \left(4 \bar{a}_2^4
  \bar{B}_{2223}-\right.\right.\right.\\
  &\left.\left.\left. \bar{a}_2^2 \bar{C}_{2223}+\bar{a}_3^2
  \bar{C}_{2233}\right)-4 \left(\bar{a}_2^2+1\right) \bar{B}_{12}\right)+20
  \bar{a}_2^4 \bar{B}_{12}^2+\bar{a}_3^2 \left(\bar{a}_3^2-\bar{a}_2^2\right)
  \bar{A}_3 \left(\bar{a}_2^2 \left(3\left(\bar{C}_{1223}+\bar{C}_{2223}\right)+
  5 \bar{C}_{2233}\right)+\right.\right.\\
  &\left.\left.\bar{a}_3^2 \left(-3   
  \left(\bar{C}_{1233}+\bar{C}_{2233}\right)+20 \bar{a}_3^2 \bar{B}_{2333}-5
  \bar{C}_{2333}\right)-12 \bar{a}_2^4 \bar{B}_{2223}\right)+\frac{2
  \left(w_1+w_2\right) \left(2 \bar{a}_2^2+1\right) \bar{a}_2^3 \bar{\Omega
  }}{\bar{a}_2^2-1}\right)
  \end{split}
  \end{flalign}
  \begin{flalign*}
  \begin{split}
  b_3&=-\left(-2 \bar{A}_1 \left(2 \bar{B}_{113}-\bar{C}_{113}+4 \bar{a}_2^2
  \bar{B}_{12}+\bar{a}_3^2 \left(\bar{C}_{133}+3 \bar{A}_3\right)\right)+4
  \bar{A}_{\emptyset } \left(-\bar{A}_1+2 \bar{B}_{12}+\bar{a}_3^2
  \bar{A}_3\right)+3 \bar{a}_3^2 \bar{B}_{11} \bar{A}_3+\right.\\
  &\left.\bar{B}_{12} \left(\bar{a}_3^2 \left(-4 \bar{B}_{123}-8 \bar{B}_{13}+
  4 \bar{C}_{233}+3 \bar{A}_3\right)-4 \bar{a}_2^2 \left(-2 \bar{B}_{113}+
  \bar{C}_{113}-\bar{a}_3^2 \left(2 \bar{B}_{123}+4 \bar{A}_{13}+
  3 \bar{B}_{13}-2
  \bar{A}_3\right)+\right.\right.\right.\\
  &\left.\left.\left.\bar{a}_3^4 \bar{B}_{133}\right)\right)+\bar{a}_3^2
  \bar{A}_3 \left(-3 \left(\bar{a}_3^2+2\right) \bar{B}_{113}+\bar{a}_3^2
  \left(-\left(3 \bar{B}_{123}+3 \bar{C}_{133}+5 \bar{a}_3^2 \left(2
  \left(\bar{a}_3^2-1\right) \bar{B}_{1333}-\bar{C}_{1333}-2
  \bar{B}_{333}\right)+\right.\right.\right.\right.\\
  &\left.\left.\left.\left.5 \bar{C}_{1333}+3 \bar{C}_{233}+5
  \bar{B}_{33}\right)\right)+5 \bar{B}_{13}\right)-2 \bar{a}_2^2 \bar{A}_2
  \left(3\bar{B}_{12}+\bar{a}_3^2\left(\bar{B}_{123}+
  \bar{C}_{233}\right)\right)-4 \bar{B}_{12}^2+6 \bar{a}_3^4
  \bar{A}_3^2+\left(w_2-w_1\right) \bar{a}_2 \bar{\Omega }\right)\\
  b_4&=-\frac{2}{a_2^2} \left(2 \left(\bar{a}_2^2-\bar{a}_3^2\right) \bar{A}_1
  \left(2 \bar{B}_{1123}-\bar{C}_{1123}+\bar{a}_3^2 \bar{C}_{1233}\right)+4
  \bar{B}_{12} \left(\bar{a}_2^2 \left(6 \bar{B}_{112}+\bar{a}_3^2 \left(-3
  \left(\bar{C}_{1123}+\bar{C}_{1223}\right)+\right.\right.\right.\right.\\
  &\left.\left.\left.\left. \bar{a}_3^2 \bar{C}_{1233}+ 2 \bar{B}_{13}\right)+3
  \left(\bar{C}_{1223}+2 \bar{B}_{23}\right)\right)+\bar{a}_2^4 \left(3
  \bar{C}_{1123}+6 \bar{B}_{122}+6 \bar{B}_{123}-\bar{a}_3^2
  \bar{C}_{1233}\right)+ \right.\right.\\
  &\left.\left.\bar{a}_3^2 \left(\left(\bar{a}_3^2-1\right) \bar{C}_{1233}+2
  \bar{B}_{23}\right)\right)+\bar{a}_3^2 \bar{A}_3 \left(\bar{a}_3^2 \left(3
  \left(\bar{C}_{1123}+\bar{C}_{1223}\right)-6 \bar{B}_{1123}-\bar{a}_3^2
  \left(3 \left(\bar{C}_{1233}+\bar{C}_{2233}\right)+10
  \bar{B}_{1333}+\right.\right.\right.\right.\\
  &\left.\left.\left.\left. 5 \bar{C}_{2333}\right)+5 \bar{C}_{1233}+10
  \bar{a}_3^4 \left(\bar{B}_{1333}+\bar{B}_{2333}\right)\right)+\bar{a}_2^2
  \left(-3 \left(\bar{C}_{1123}+\bar{C}_{1223}\right)+6
  \bar{B}_{1123}+\bar{a}_3^2 \left(3\left(\bar{C}_{1233}+
  \bar{C}_{2233}\right)-\right.\right.\right.\right.\\
  &\left.\left.\left.\left.10 \bar{a}_3^2 \bar{B}_{2333}+5
  \bar{C}_{2333}\right)-5 \bar{C}_{1233}\right)-6 \left(\bar{a}_3^2-1\right)
  \bar{a}_2^4 \bar{B}_{1223}\right)+12 \bar{a}_2^2  \bar{B}_{12}^2+2 \bar{a}_2^2
  \bar{A}_2 \left(-2 \left(\bar{a}_3^2-1\right) \bar{a}_2^4 \bar{B}_{1223}+
  \right.\right.\\
  &\left.\left. \left(\bar{a}_3^2-\bar{a}_2^2\right) \bar{C}_{1223}+\bar{a}_3^2
  \left(\bar{a}_2^2-\bar{a}_3^2\right) \bar{C}_{2233}\right)+\frac{6
  \left(w_1+w_2\right) \bar{a}_2^3 \bar{\Omega }}{\bar{a}_2^2-1}\right)\\
  b_5&=-6 \left(2 \left(\bar{a}_3^2-1\right) \bar{A}_1 \left(-5
  \bar{B}_{1113}+\bar{B}_{113}-\bar{a}_3^2   
  \bar{C}_{1133}\right)-24 \bar{a}_2^2 \bar{C}_{111}
  \bar{B}_{12}-20 \bar{a}_2^2 \bar{B}_{111} \bar{B}_{12}+\right.\\
  &\left. 4 \bar{B}_{12} \left(\bar{a}_2^2 \left(4 \left(\bar{a}_3^2-1\right)
  \bar{B}_{1113}+6 \bar{B}_{112}-2 \bar{a}_3^2 \bar{B}_{113}+6
  \bar{B}_{122}+4 \bar{B}_{123}+4 \bar{B}_{13}+\bar{a}_3^4 \bar{B}_{133}+2
  \bar{B}_{23}\right)+\bar{B}_{112}+\right.\right.\\
  &\left.\left.2 \bar{a}_3^2 \left(\bar{B}_{123}+2 \bar{C}_{123}\right)+2
  \bar{a}_2^4 \bar{B}_{123}+4 \bar{B}_{23}+\bar{a}_3^4 \bar{B}_{233}\right)+
  \bar{a}_3^2 \left(\bar{a}_3^2-1\right) \bar{A}_3 \left(3
  \left(\bar{C}_{1113}+\bar{C}_{1123} \right)-12 \bar{B}_{1113}-\right.\right.\\
  &\left.\left.\bar{a}_3^2 \left(3 \left(\bar{C}_{1133}+\bar{C}_{1233}\right)+5
  \bar{C}_{1333}\right)+5 \bar{C}_{1133}+20 \bar{a}_3^4 \bar{B}_{1333}\right)+2
  \left(\bar{a}_3^2-1\right) \bar{a}_2^2 \bar{A}_2
  \left(\bar{C}_{1123}-\bar{a}_3^2 \bar{C}_{1233}\right)+\right.\\
  &\left.4 \left(2 \bar{a}_2^2+3\right) \bar{B}_{12}^2+\frac{2
  \left(w_1+w_2\right) \left(\bar{a}_2^2+2\right) \bar{a}_2 \bar{\Omega
  }}{\bar{a}_2^2-1}\right).
\end{split}
\end{flalign*}

\section{The limit $\ba_3\to 0$ to 1 PN order}\label{sec:limit_rod}

In \cite{Gurlebeck_2010}, we took a close look at the axisymmetric limit $(\ba_2
\to 1)$ of the PN Dedekind ellipsoids. It turns out that the PN Maclaurin
spheroids emerge in the limit but only if the PN velocity field is generalized
as in Equation \eqref{eq:ansatz}. Here we consider the opposite limit of a rod
along the $x_1$-axis, i.e., $\ba_2 \to 0$, cf. Section \ref{sec:limiting_cases}.
We begin by deriving conditions that arise from the behavior of the metric
functions in this limit, where it will be important to treat $\mu$ and $a_1$ as
functions of $\ba_2$. We then examine the surface and conclude that the only
acceptable solution is a member of the Weyl class, i.e., an axially symmetric
and static spacetime. The matter content collapses to a singularity along the
axis and the limiting spacetimes contain the Levi-Civita spacetime and the Curzon-Chazy
particle in the special cases in which the rod has infinite length
($a_1\to\infty$) or zero length ($a_1 \to 0$), see e.g.\ \cite{Griffiths_2009}.
In order to include the Levi-Civita metric in the subsequent derivations, we
shall divide up the ellipsoid into slices defined by $x^1=x^1_0$ and with a
thickness $\delta x^1$, which we denote by $\mathcal S\left(\delta
x^1,x^1_0\right)$.

Let us start by looking at $\Phi$, which is determined by the Poisson equation
\eqref{eq:1 PNPhi}. The inhomogeneity is a sum of (Newtonian) kinetic, inner and
potential energy densities. Beginning with the kinetic energy contained in
$\mathcal S$, we find
\begin{align*}
\begin{split}
&\int\limits_{\mathcal S} \frac{\mu}{2} \mathbf v^2 d^3 x  \propto
a_1^4\ba_2 \ba_3 \delta x^1  \mu^2 \bB_{1
2}\left(1-\left(\frac{x^1_0}{a_1}\right)^2\right) 
\left(1+(4\ba_2^2-1)\left(\frac{x^1_0}{a_1}\right)^2+O\left(\frac{\delta
x^1}{a_1}\right)\right).
\end{split}
\end{align*}
It turns out that the ratio of the inner to the kinetic energy tends to zero in
the limit. Furthermore, the potential energy is proportional to the kinetic one.
Thus, it suffices to derive the form of the kinetic energy density in the limit.
We choose it to be a line energy density as required for a well-defined Equation \eqref{eq:1 PNPhi}. This then implies
\begin{align}\label{eq:mass_density_limit}
 \mu=\frac{e_{\mathrm{max}}^{\frac 12}}{\pi \sqrt{G} a_1^2 \ba_3^2
 (-\ln\ba_3)^{\frac 1 2}},
\end{align}
whereby $e_{\mathrm{max}}$ is an arbitrary constant and the logarithmic term
comes from the expansion of $\bar B_{12}$. This yields a source for the
potential $\Phi$ of the form
\begin{align}\label{eq:linedensityPhi}
  3e_{\mathrm{max}} G\left(1-\left(\frac{x^1}{a_1}\right)^2
  \right)^2\delta(x^2)\delta(x^3)\Theta(a_1^2-x_1^2),
\end{align} 
where $\delta(x)$ and $\Theta(x)$ denote the Dirac delta distribution and the
Heaviside step function, respectively. In cylindrical coordinates ($x^2=\rho
\cos\varphi$ and $x^3=\rho \sin\varphi$) the solution to Equation \eqref{eq:1
PNPhi} reads
\begin{align}\label{eq:Phi2}
\begin{split}
  \Phi=&\frac{e_{\mathrm{max}} G}{8 a_1^4} \left(\vphantom{\ln
  \left(\frac{\sqrt{\left(a_1-x^1\right){}^2+\rho ^2}+
  a_1-x^1}{\sqrt{\left(a_1+x^1\right){}^2+\rho^2}-a_1-x^1}\right)}
  \sqrt{\left(a_1+x^1\right){}^2+\rho ^2} \left(-9 a_1 \rho ^2+ 58 a_1^2 x^1+26
  a_1 \left(x^1\right)^2-18 a_1^3+55 \rho ^2 x^1-50
 \left(x^1\right)^3\right)+\right.\\
  &\left.\sqrt{\left(a_1-x^1\right){}^2+\rho ^2} \left(-9 a_1 \rho ^2-58 a_1^2
  x^1+26 a_1 \left(x^1\right)^2-18 a_1^3-55 \rho ^2 x^1+50
  \left(x^1\right)^3\right)+\right.\\
  &\left.3 \left(8 \rho ^2 \left(a_1^2-3 \left(x^1\right)^2\right)+8
  \left(a_1^2-\left(x^1\right)^2\right){}^2+3 \rho ^4\right)\ln
  \left(\frac{\sqrt{\left(a_1-x^1\right){}^2+\rho
  ^2}+a_1-x^1}{\sqrt{\left(a_1+x^1\right){}^2+\rho^2}-a_1-x^1}\right)\right).
\end{split}
\end{align}

Now consider the potentials $U_\alpha$. We remind the reader that $U_3$ vanishes
for all $\ba_2$. We prove that $U_{1}$ is also zero in the exterior in the limit
by looking at its multipoles. The inhomogeneity in Equation \eqref{eq:Uvec} is
proportional to the Newtonian linear momentum density in the $x^1$-direction.
Its integral over a slice $\mathcal S\left(\delta x^1,x^1_0\right)$ vanishes
because of the antisymmetry of $v^{(0)1}$. The integrals over the halves of the
slice with $x^2>0$ and $x^2<0$ are given to leading order in $\ba_3$ by
\begin{align}
 P^1_{\pm}=&\mp 2\frac{e_{\mathrm{max}}^{\frac 3 4}}{\pi G^{\frac 1 4}} \delta
 x^1 \left(1-\left(\frac{ x^1_0}{a_1}\right)^2+O\left(\frac{\delta
 x^1}{a_1}\right)\right) \left(-\ln \bar{a}_3\right){}^{-\frac 1 4},
\end{align}
which evidently tend to zero for $\ba\to 0$. An arbitrary multipole moment for,
e.g.\, the density $\mu v^{(0)1}$, again to leading order, is then bounded by
\begin{align}\label{eq:PPlusMinus}
  \int\limits_{\mathcal S} \left|\mu v^{(0) 1} 
  {\left(x^1\right)}^i{\left(x^2\right)}^j {\left(x^3\right)}^k\right| \,
  d^3x \leq {\left(a_1\right)}^i{\left(a_2\right)}^j
  {\left(a_3\right)}^k \int\limits_{\mathcal S} \left|\mu v^{(0)1}\right| \,
  d^3x = {\left(a_1\right)}^i{\left(a_2\right)}^j
  {\left(a_3\right)}^k\left(|P^1_+| + |P^1_-| \right),
\end{align}
which all tend to zero by virtue of the preceding equation. This proves that
$U_1$ vanishes for $\ba_2\to 0$ in the exterior. This holds via
corresponding arguments for $U_2$, too. Hence, the time-like
Killing vector is hypersurface orthogonal and the spacetime is static in the limit.

It remains to analyze $\delta U$ and the PN surface. With the mass density given
in Equation \eqref{eq:mass_density_limit} one can show that $U$ tends to zero in
the exterior analogously to $U_\alpha$. It, thus, suffices to consider the
solution to the Poisson equation for $U'$ from Equation \eqref{eq:PoissonUprime}. The 1
PN surface is still defined by the condition of vanishing pressure at the
surface. A necessary condition for the inhomogeneity in Equation
\eqref{eq:PoissonUprime} to have a well-defined limit in a distributional sense
is that the mass $\delta M$ contained in a slice $\mathcal S$ is well-defined to
1 PN order:
\begin{align}\label{eq:deltaM}
  \delta M=\int\limits_{x^1_0}^{x^1_0+\delta x^1}\int\limits_{0}^{2\pi}
  \int\limits_{0}^{\rho_s}\mu \,\rho d \rho\, d \varphi\, d
  x^1=\frac{\mu}{2}\int\limits_{x^1_0}^{x^1_0+\delta
  x^1}\int\limits_{0}^{2\pi}\rho_s^2d \rho\, d \varphi\, d x^1,
\end{align}
where $\rho_s$ denotes the 1 PN surface \eqref{eq:surfacePN} in cylindrical
coordinates. Using our standard notation for PN terms, the 1 PN order of the
relevant term of the integrand is
\begin{align}\label{eq:integrand_for_delta_M}
  \left(\rho_s(x^1,\varphi)\right)^2=\left(\rho^{(0)}(x^1,\varphi)\right)^2+
  2\rho^{(0)}(x^1,\varphi)\rho^{(2)}(x^1,\varphi)c^{-2}.
\end{align}
The integral over the Newtonian contribution vanishes because of Equation
\eqref{eq:mass_density_limit}. The second term can be written out
explicitly using Equation \eqref{eq:ansatz} and an expansion
in terms of $\ba_3$. The expansions of $w_j$ and $S_i$ will be denoted by
\begin{align}
  w_j&=w_{j 0}(\ln\ba_3) + w_{j 2}(\ln\ba_3) \ba_3^2 + O(\ba_3^3), \quad
  j\in\{1,2\},\\
  \begin{split}
    S_i&=
    \begin{cases}
      S_{i 0}(w_{jk},\ln\ba_3) + S_{i 2}(w_{jk},\ln\ba_3) \ba_3^2+ O(\ba_3^3),
      \quad i\in\{1,2,3\}\\
      S_{i 0}(w_{jk},\ln\ba_3)\ba_3^{-2} + S_{i 2}(w_{jk},\ln\ba_3) +
      O(\ba_3^3), \quad i\in\{4,5\}.
    \end{cases}
  \end{split}
\end{align}
The expression \eqref{eq:integrand_for_delta_M} diverges like $\ba_3^{-2}$ for
$\ba_3\to 0$ in general. Hence, $\delta M$ becomes singular in this limit as
well, see Equation \eqref{eq:deltaM}. These diverging terms vanish only if we
have:
\begin{align}\label{eq:leading_order_condition}
  S_{50}=4S_{10}, \quad S_{50}=-\frac{4}{3}S_{30}.
\end{align}
However, the $S_{i0}$ must also be consistent with Equation \eqref{eq:surface_condition}
to leading order, which provides the further conditions
\begin{align}\label{eq:leading_order_S}
\begin{split}
  0=&6 S_{10} + 12 S_{20} - 10 S_{40} -5 S_{50},\\
  0=& 4 S_{40} +  S_{50},\\
  0=& S_{10} \left(\frac{25}{12} + \ln\frac{\ba_3}{2}\right) + \frac{S_{20}}{6}
  - \frac{S_{30}}{3} \left(\frac{31}{12} +  \ln\frac{\ba_3}{2}\right) -
  \frac{S_{40}}{36}   -    \frac{S_{50}}{2} \left(\frac{197}{72}  + 
  \ln\frac{\ba_3}{2}\right) +   \frac{w_{10}- w_{20}}{6} \sqrt{- \frac{3}{2} -
  \ln\frac{\ba_3}{2}},\\
  0=& 6 S_{30} - 10 S_{40} - 11 S_{50},\\
  0=& 4\left(S_{10}-S_{20}-\frac{1}{2} \right)\left(\frac{3}{2} +
  \ln\frac{\ba_3}{2}\right) - 3 S_{30} \left(\frac{109}{54} +
  \ln\frac{\ba_3}{2}\right) - 2 S_{40} \left(\frac{73}{48} + 
  \ln\frac{\ba_3}{2}\right) - \frac{5}{4} S_{50} \left(\frac{23}{8} + \ln\frac{
  \ba_3}{2}\right) +\\  & (w_{10} + w_{20}) \sqrt{-\frac{3}{2} - \ln
  \frac{\ba_3}{2}} .
\end{split}
\end{align}
Together with Equations \eqref{eq:leading_order_condition} these equations imply
that
\begin{align}\label{eq:S_i_zero}
\begin{split}
  S_{i0}=0\quad i\in\{1,2,\ldots,5\}, \quad 
  w_{10}=w_{20}=\sqrt{-\frac 3 2 - \log \frac{\ba_3}{2}}.
\end{split}
\end{align}

Although these conditions are necessary they are not sufficient. The next to
leading order in $\ba_3$ in $\delta M$ diverges in general logarithmically. To
choose the parameters $w_{i2}$ so that these terms vanish, we solve first the
system of equations originating from Equation \eqref{eq:surface_condition}, which reads
to this order
\begin{align}\label{eq:next_to_leading_order_eq_S}
\begin{split}
  0=&6 S_{12}+12 S_{22}-10 S_{42}-5 S_{52}+16  w_{10}^2\\
  0=&4S_{42} + S_{52} + 4 w_{10}^2\\
  0=&S_{12}\left(12w_{10}^2-7\right)-2 S_{22}-S_{32}\left(\frac{13}{3}-4
  w_{10}^2\right)+\frac{1}{3}S_{42}+S_{52}\left(\frac{89}{12}-6
  w_{10}^2\right)-2w_{10}\left(w_{12}-w_{22}\right)-16 w_{10}^4+\\
  &\frac{8}{3}w_{10}^2+5=0\\
  0=&6 S_{32}-10 S_{42}-11 S_{52}+40 w_{10}^2\\
  0=& 96 w_{10}^2\left(S_{12} - S_{22} \right) +  S_{32} \left(\frac{112}{3} -
  72 w_{10}^2\right) + S_{42} \left(1 - 48 w_{10}^2\right) + S_{52} \left(
  \frac{165}{4} - 30 w_{10}^2\right) + 24 w_{10} \left(w_{12} + w_{22}\right)
  -\\
  & 144 w_{10}^4 - 100 w_{10}^2 + 5.
\end{split}
\end{align}
This can easily be solved and the solution, which depends on $w_{i2}$, can be
inserted in $\delta M$. Now, the vanishing of the diverging terms requires the
following behavior of the free parameters $w_{i2}$:
\begin{align}\label{eq:wi2}
\begin{split}
  w_{12} &= -\frac{21}{2} (-\log \ba_3)^{\frac{3}{2}}+C_1 (-\log
  \ba_3)^{\frac{1}{2}}+C_2+O\left((-\log \ba_3)^{-\frac 12}\right),\\
  w_{22} &=C_3 (-\log \ba_3)^{\frac{3}{2}}+C_4 (-\log \ba_3) +C_5 (-\log
  \ba_3)^{\frac 12}+C_6+O\left((-\log \ba_3)^{-\frac{1}{2}}\right)
\end{split}
\end{align}
with the free constants $C_{i}$. 

The constants $C_1,~C_2,~C_3$ and $C_4$ in Equation \eqref{eq:wi2} govern the
length of the rod $\Delta x_1$ in the limit $\ba_3\to 0$, which still diverges
in general. The choice
\begin{align}\label{eq:rod_finite_length}
  C_3=\frac{1}{134} (-16715 + 9072 \log 2 +C_1)
\end{align}
ensures a finite length, which reads then
\begin{align}
  (\Delta x_1)^2=a_1^2 - \frac{a_1^2 (e_{\mathrm{max}} G)^{\frac 12}}{15 c^2} (288 C_2 -
  67 C_4).
\end{align}
With an argument like in Equation \eqref{eq:PPlusMinus}, one can show that all
multipole moments converge with the choices \eqref{eq:wi2} and
\eqref{eq:rod_finite_length} and that $\delta U$ is well-defined; the
inhomogeneity in Equation \eqref{eq:Uvec} tends to a line mass density,
which is a polynomial in $x^1$ to order 4:
\begin{align}\label{eq:linedensityUprime}
  \frac{3 G e_{\mathrm{max}}}{134 a_1^4 c^2}\left(-455 + 252 \log 2 +  16
  C_1\right)\left( a_1^4 - 6 a_1^2 x_1^2 + 5 x_1^4\right)
  \delta(x^2)\delta(x^3)\Theta(a_1^2-x_1^2).
\end{align}
This is already of order $c^{-2}$. Hence, it is sufficient to take the Newtonian
length of the rod into account.
The function $\delta U$ can easily be calculated for the line density
\eqref{eq:linedensityUprime} and is given by
\begin{align}
\begin{split}
  \delta U=&-\frac{e_{\mathrm{max}} G}{1072 a_1^4}  (-455 + 252 \log 2 + 16 C_1)
  \left[N_{+}+N_{-} +  \log\left(\frac{-a_1 - x_1 + ((a_1 + x_1)^2 +
  \rho^2)^{\frac 12}}{ a_1 - x_1 + ((a_1 - x_1)^2 + \rho^2)^{\frac 12}}\right)
  \times\right.\\
  &\left. (24 a_1^4 - 144a_1^2   x_1^2 + 120 x_1^4 +  72 a_1^2 \rho^2 - 360
  x_1^2 \rho^2 + 45 \rho^4) \vphantom{ \log\left(\frac{-a_1 - x_1 +  ((a_1 +
  x_1)^2 + \rho^2)^{\frac 12}}{ a_1 - x_1 + ((a_1 - x_1)^2 + \rho^2)^{\frac
  12}}\right)} \right],\\
  N_{\pm}=&((a_1 \pm x_1)^2 + \rho^2)^{\frac 12} \left( 42 a_1^3 \mp 146 a_1^2
  x_1 - 130 a_1 x_1^2 \pm 250 x_1^3 + 45 a_1 \rho^2 \mp 275 x_1 \rho^2\right).
\end{split}
\end{align}

With $\Phi$, $\delta U$, $U_\alpha$, and $U$ all metric functions are determined.
The metric describes an axially symmetric and static vacuum and, therefore,
belongs to the Weyl class. The matter region degenerates to a rod along the
$x_1$-axis and is described by a singularity. The singular behavior can be read
off the Kretschmann scalar and is given to leading order in $\rho$ by
\begin{align}
  R^{a b c d}R_{abcd}\sim \frac{1}{\rho^{4}(\log \frac{\rho}{a_1})^2}.
\end{align}

The class of Weyl metrics includes such prominent members as the Lanczos metric
and the Curzon-Chazy metric (for an overview of these two metrics, see, e.g.\,
\cite{Griffiths_2009}). These metrics result also in some special limits of the
1 PN Dedekind ellipsoids. First the limit $a_3\to 0$ is carried out as described
above and afterward the following limits are taken: $a_1\to \infty$ in the case of the Lanczos metric and $a_1\to 0$ in case of the Curzon-Chazy particle. In the
former case, the line densities \eqref{eq:linedensityPhi} and
\eqref{eq:linedensityUprime} become
\begin{align}
  3e_{\mathrm{max}} G \delta(x^2)\delta(x^3), \quad \frac{3 G 
  e_{\mathrm{max}}}{134}\left(-455 + 252 \log 2 +  16 C_1\right) \delta(x^2)\delta(x^3),
\end{align}
which are constant along the entire $x^1$-axis. The resulting
spacetime is cylindrically symmetric. The leading order in $c^{-1}$ of
the mass parameter is vanishing, such that it can be interpreted  as an effective
gravitational mass per unit length, cf. \cite{Israel_1977,Griffiths_2009}.

In the case $a_1\to 0$, the density \eqref{eq:linedensityUprime} vanishes
and the density \eqref{eq:linedensityPhi} tends to the point density $E
\delta(x^1)\delta(x^2)\delta(x^3)$ with $E$ is the total Newtonian energy
concentrated in this point. This coincides with a 1 PN approximation to the
Curzon-Chazy solution with a parameter\footnote{We use here the same
notation as in \cite{Griffiths_2009}.} $m$ with a vanishing leading order in
$c^{-1}$.

\section{The plots of the solutions}\label{appendix:solution_plotted}

As some analytical expressions are lengthy they were at some places suppressed in
the article. In Figure \ref{Fig:parameters}, the main parameters describing the solution are plotted. 
We use for all parameters, say, $t_1$ the usual splitting
$t_1=t_{1,0}+t_{1,1} \hat w_1+t_{1,2} w_2$ with the parameterization from Section
\ref{sec:singularities}. Subsequently, the coefficients $t_{1,i}$ are depicted.
We use solid lines for $i=0$, dashed lines for $i=1$, and dot-dashed lines for $i=2$.
This parameterization has the advantage that the singularity is already removed.
The results reflecting the parameterization in Paper II are recovered if we set
$\hat w_1=-\tfrac{0.354937}{\lambda}$ and $w_2=0$, which implies $w_1=0$, cf.\
Equation \eqref{eq:sing_value_w1}. In Figure \ref{fig:surface}, three qualitatively different cases,
cf. Section \ref{sec:gravitomagnetic}, of the surfaces of the 1 PN Dedekind ellipsoids are shown.
\newcommand{\scaleimagesS}{0.4265}
\begin{figure}[htb!]
\begin{center}
\addtocounter{subfigure}{6}
\subfigure{\includegraphics[scale=\scaleimagesS]{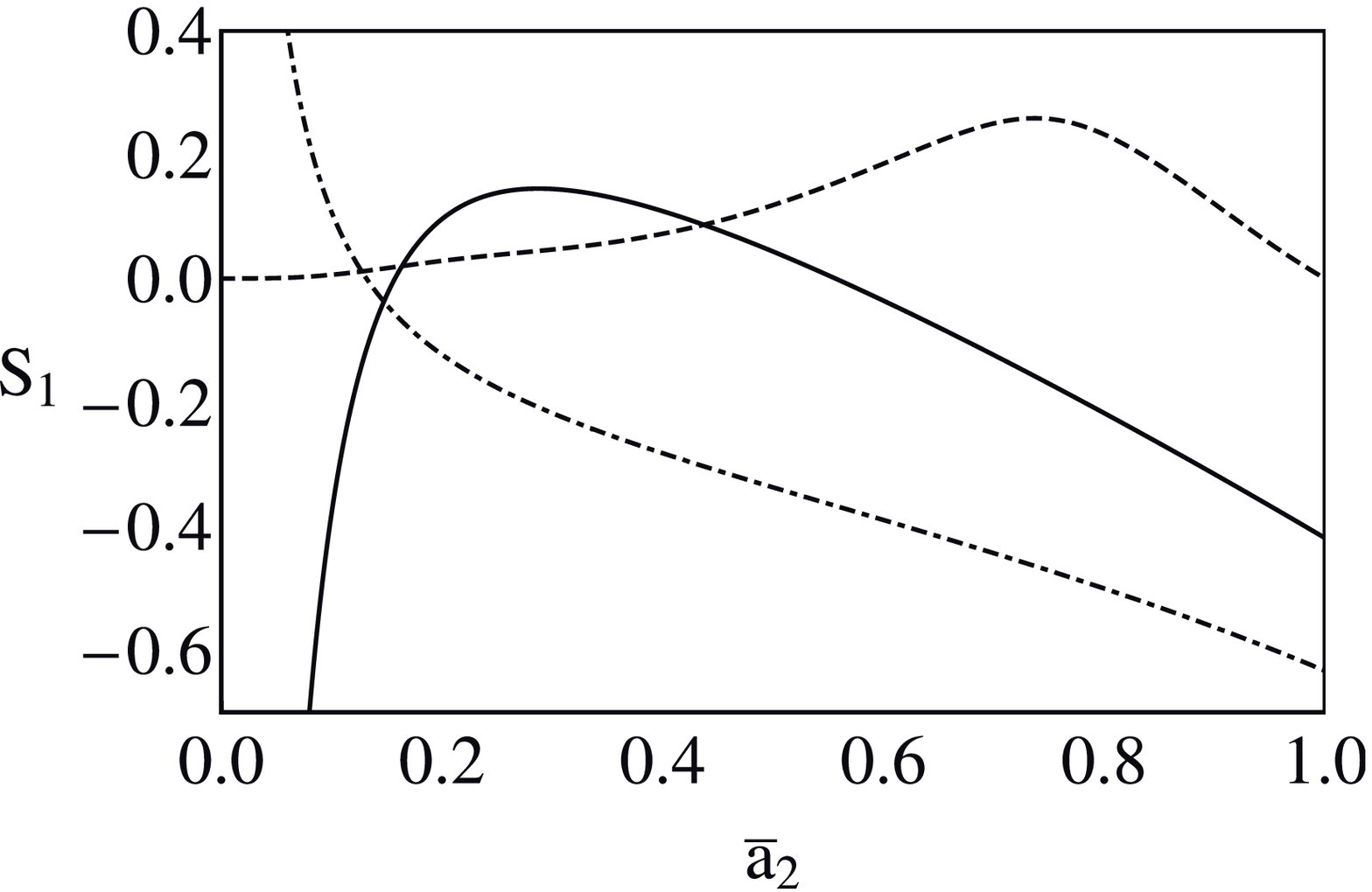}
\label{Fig:S_1}
}
\subfigure{\includegraphics[scale=\scaleimagesS]{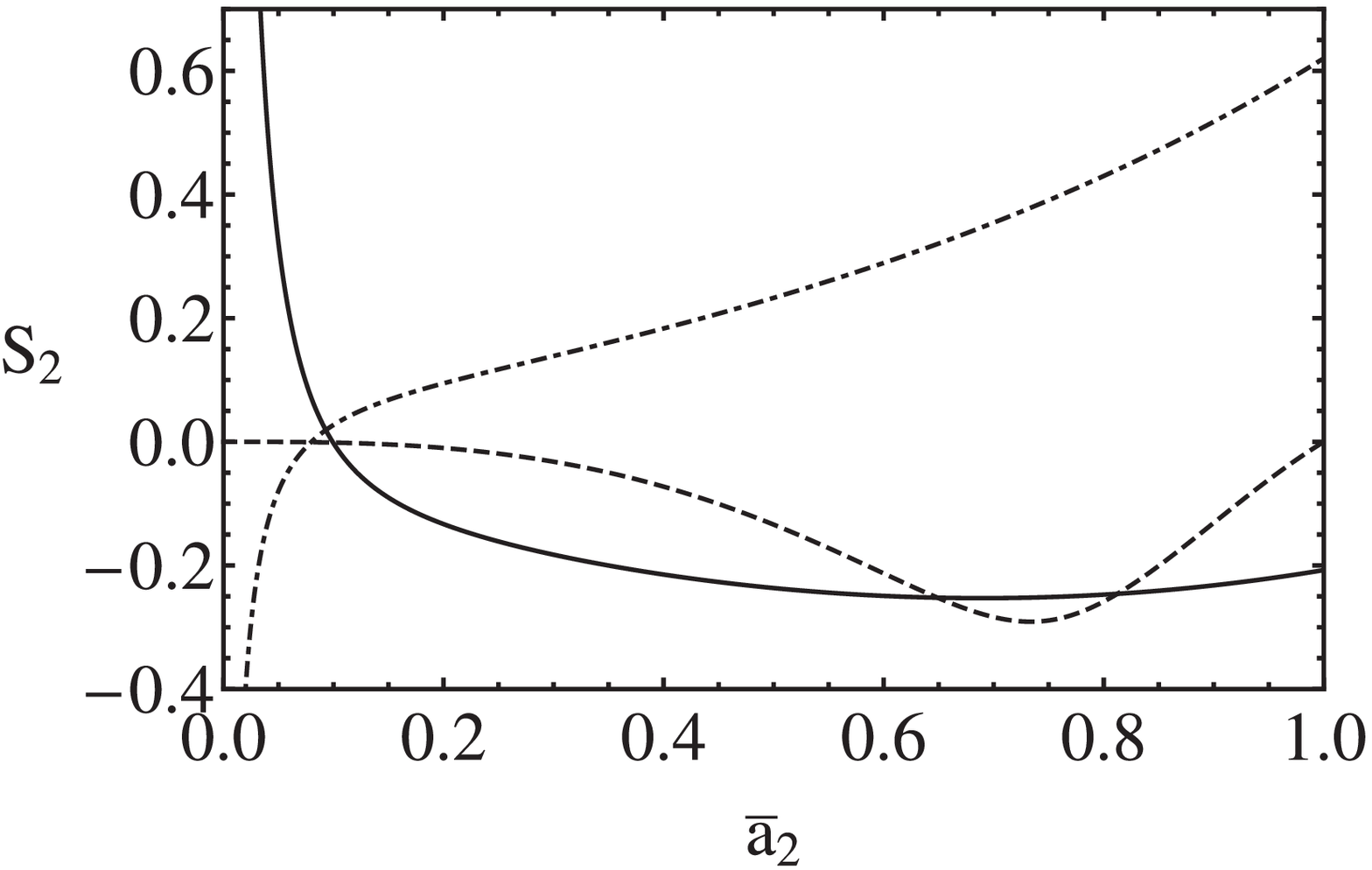}
\label{Fig:S_2}
}\\
\subfigure{\includegraphics[scale=\scaleimagesS]{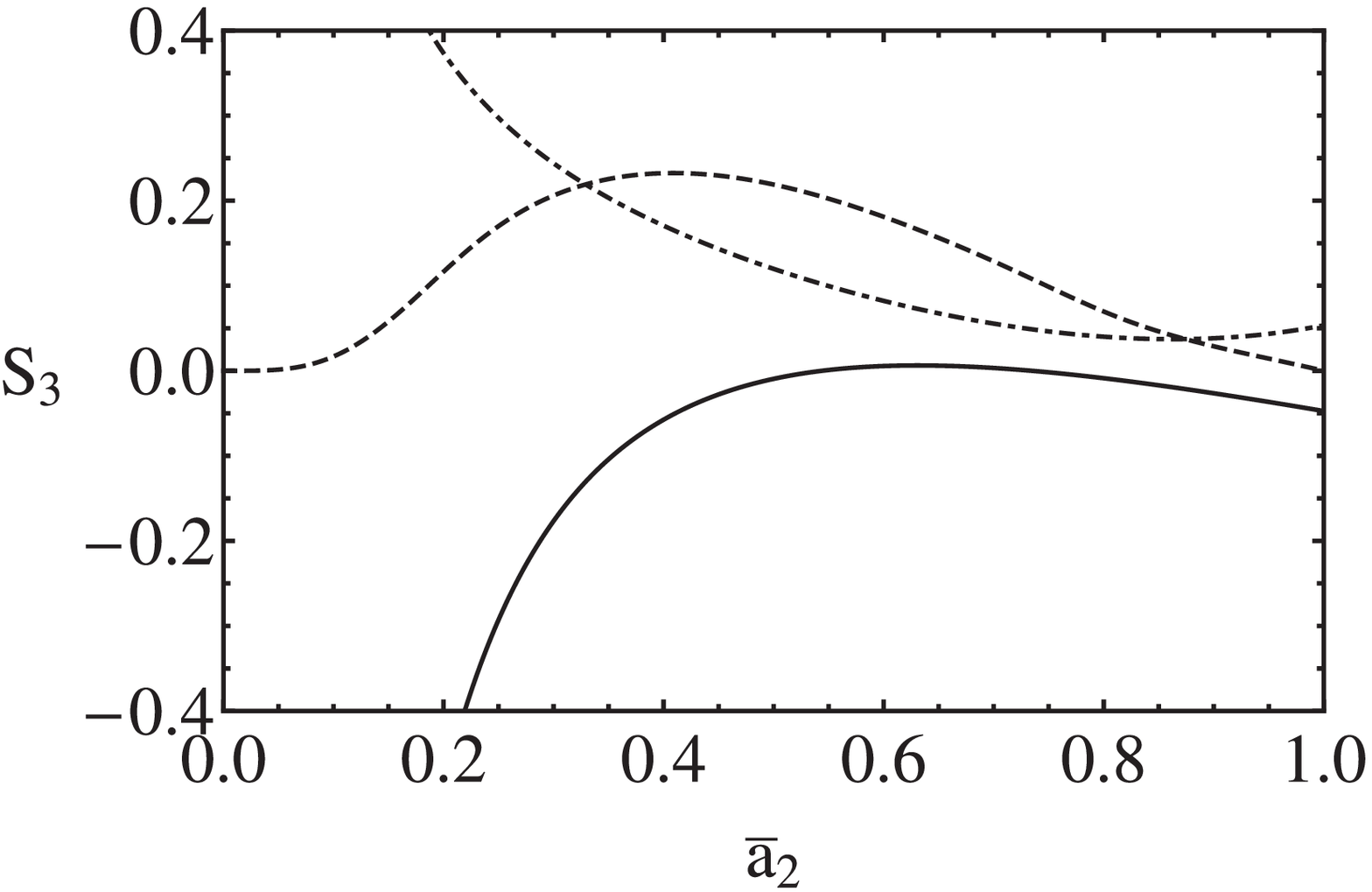}
\label{Fig:S_3}
}
\subfigure{\includegraphics[scale=\scaleimagesS]{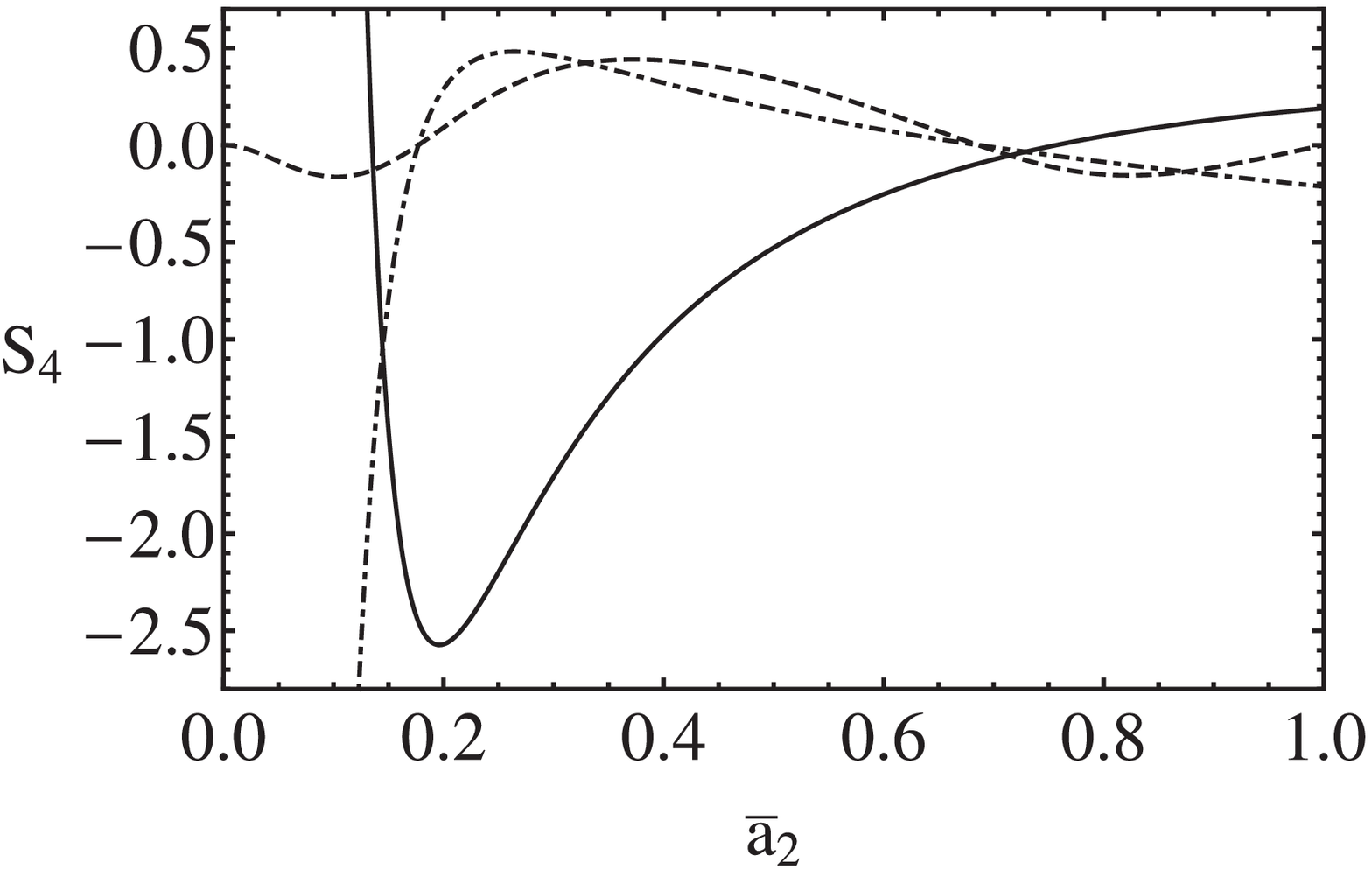}
\label{Fig:S_4}
}
\caption{The parameters $S_i,~t_i,~r_i$ and $q_i$ describing the 1 PN
corrections of the surface and the velocity field, cf.\ Equation
\eqref{eq:ansatz}.}
\label{Fig:parameters}
\end{center}
\end{figure}
\begin{figure}[htb!]
\begin{center}
\addtocounter{figure}{-1}
\subfigure{\includegraphics[scale=\scaleimagesS]{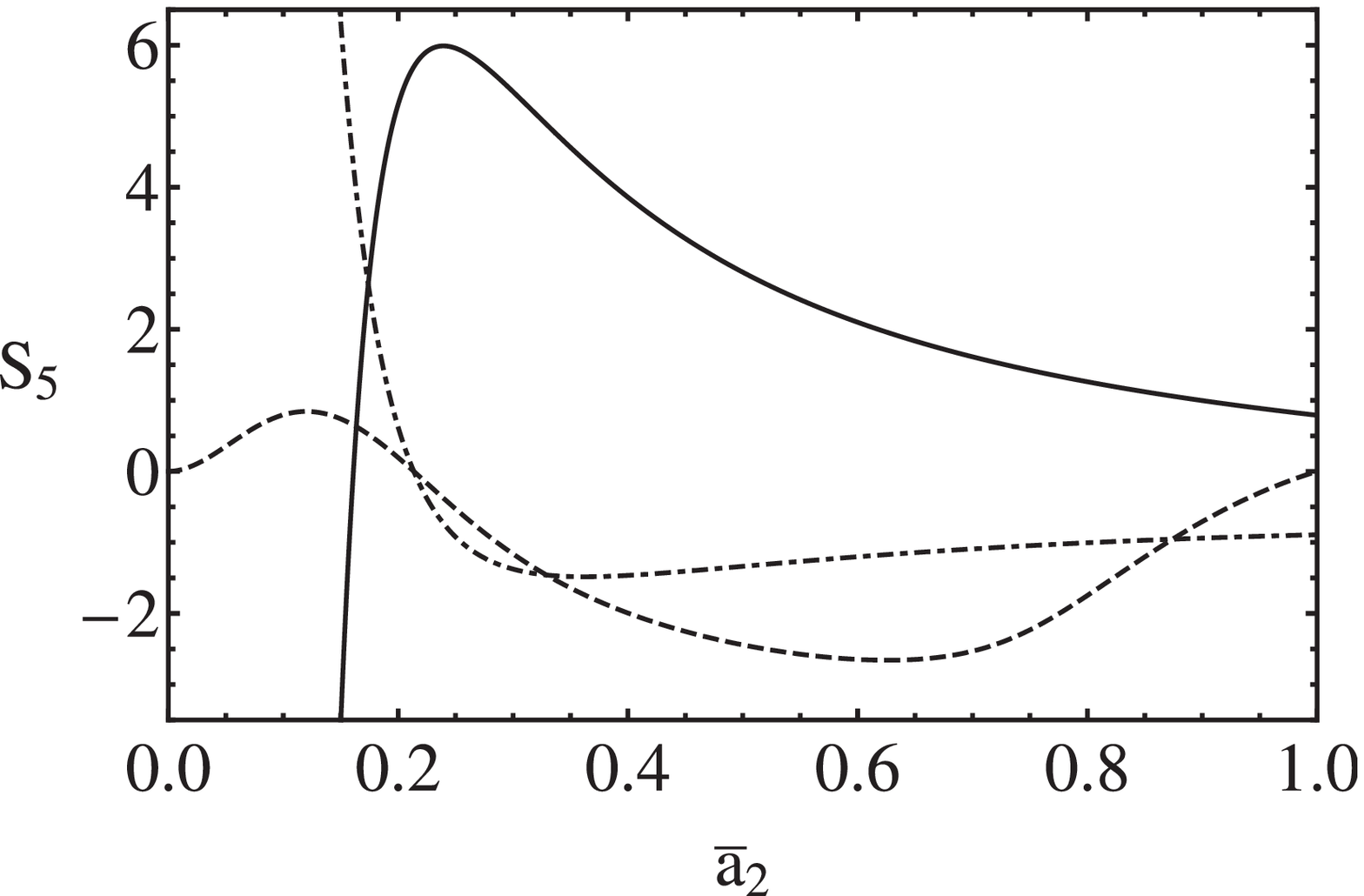}
\label{Fig:S_5}
}
\subfigure{\includegraphics[scale=\scaleimagesS]{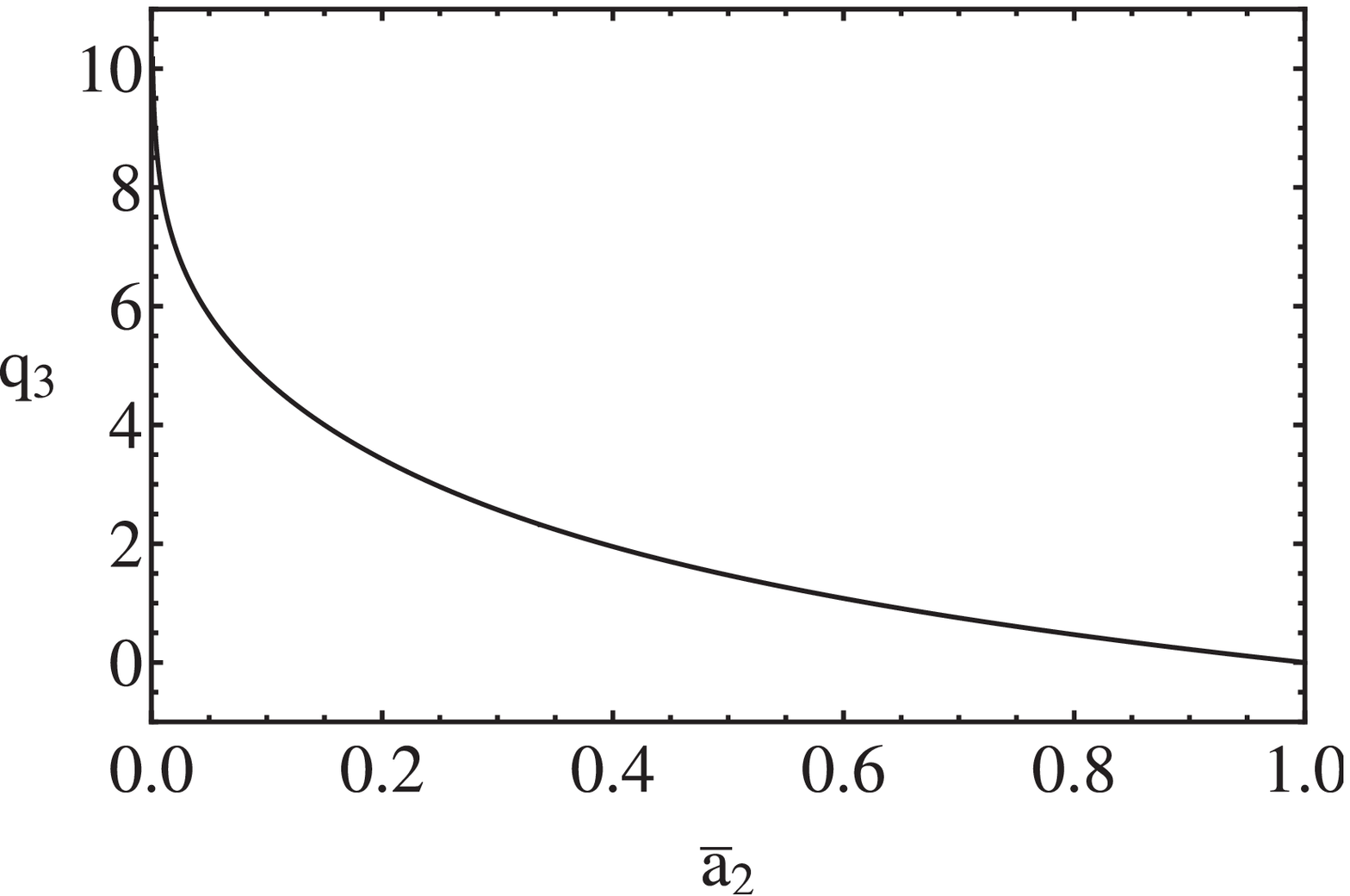}
\label{Fig:q_3}
}\\
\subfigure{\includegraphics[scale=\scaleimagesS]{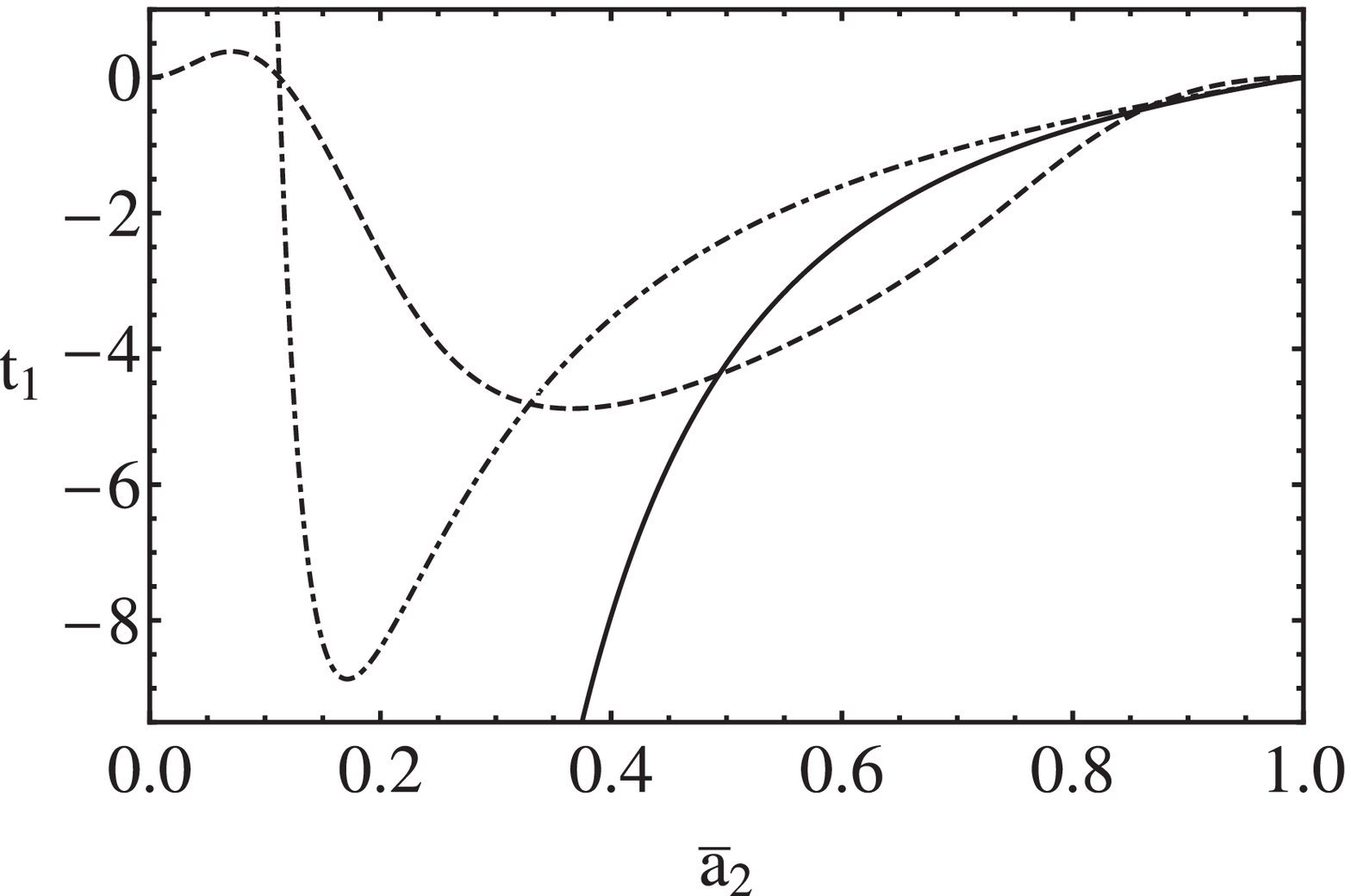}
\label{Fig:t_1}
}
\subfigure{\includegraphics[scale=\scaleimagesS]{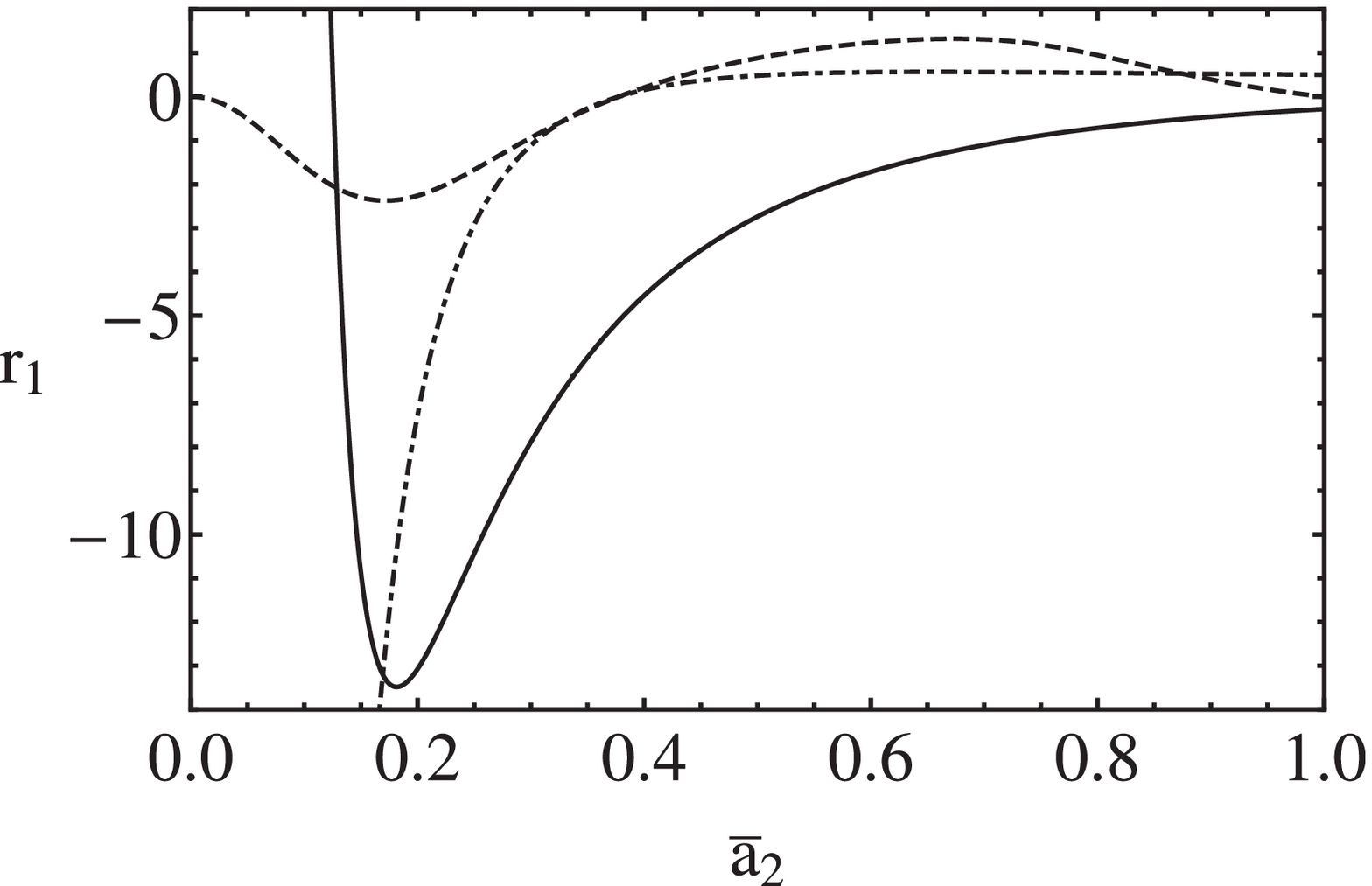}
\label{Fig:r_1}
}\\
\subfigure{\includegraphics[scale=\scaleimagesS]{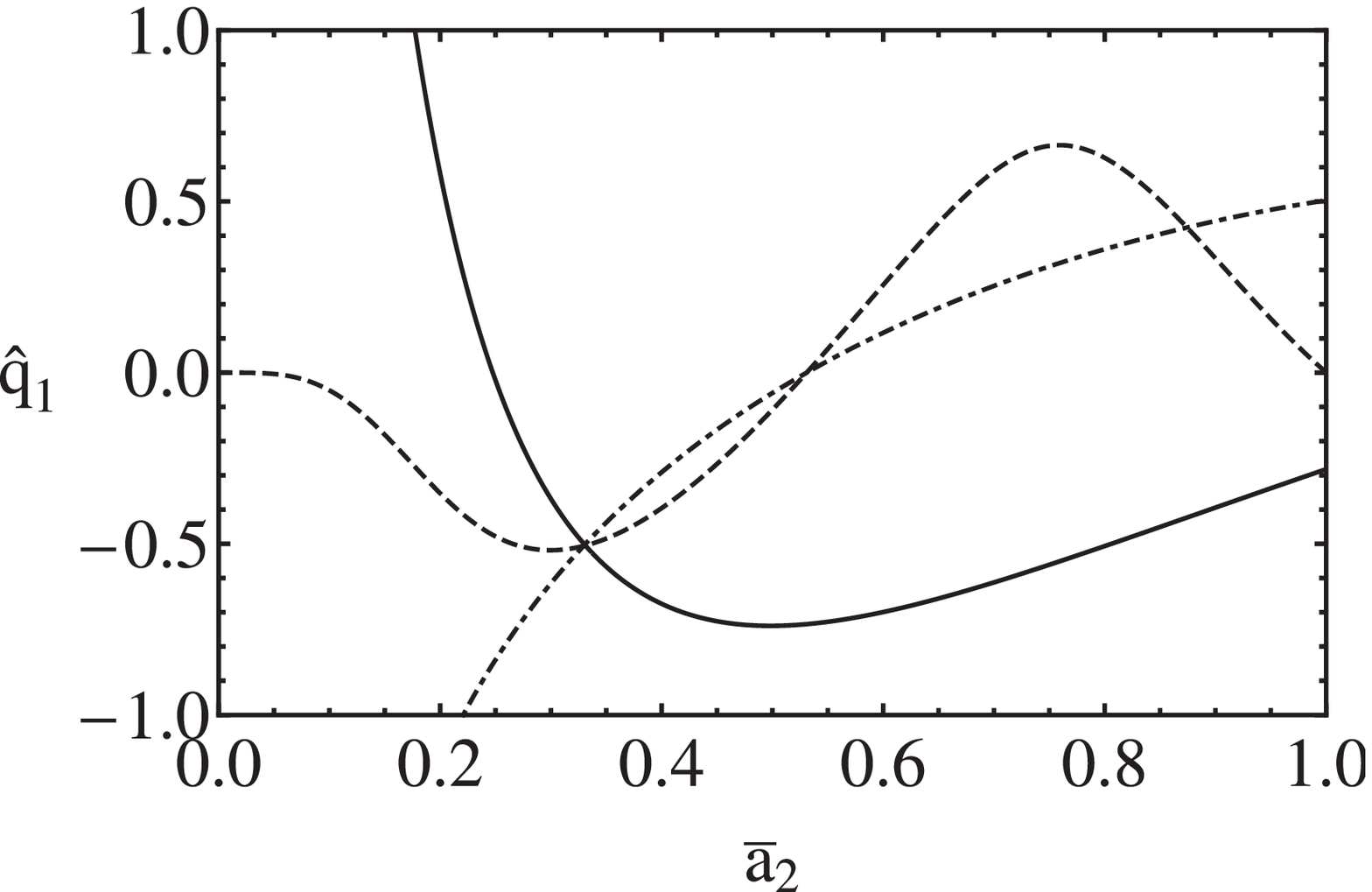}
\label{Fig:p_1}
} 
\subfigure{\includegraphics[scale=\scaleimagesS]{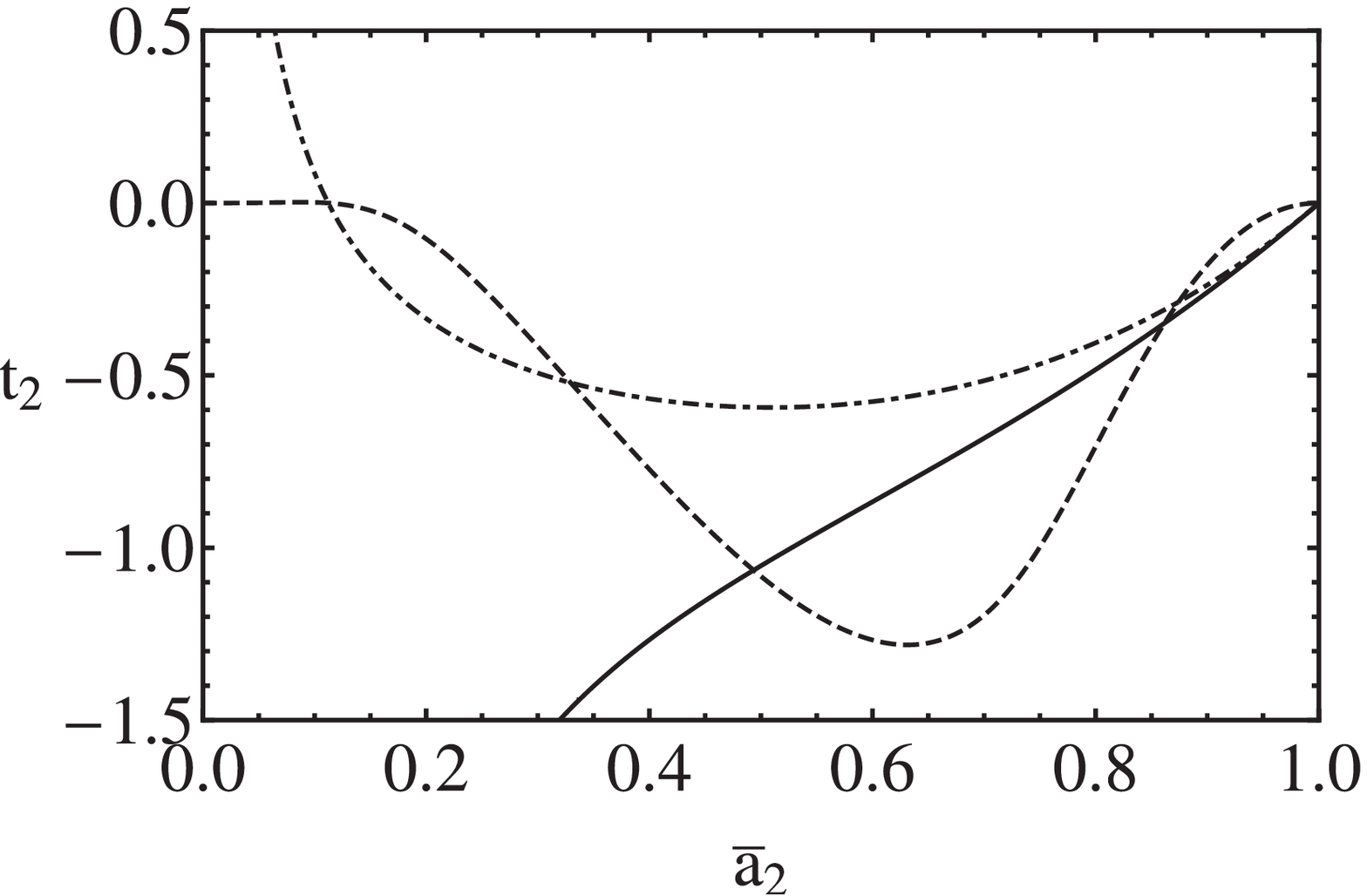}
\label{Fig:t_2}
}\\
\subfigure{\includegraphics[scale=\scaleimagesS]{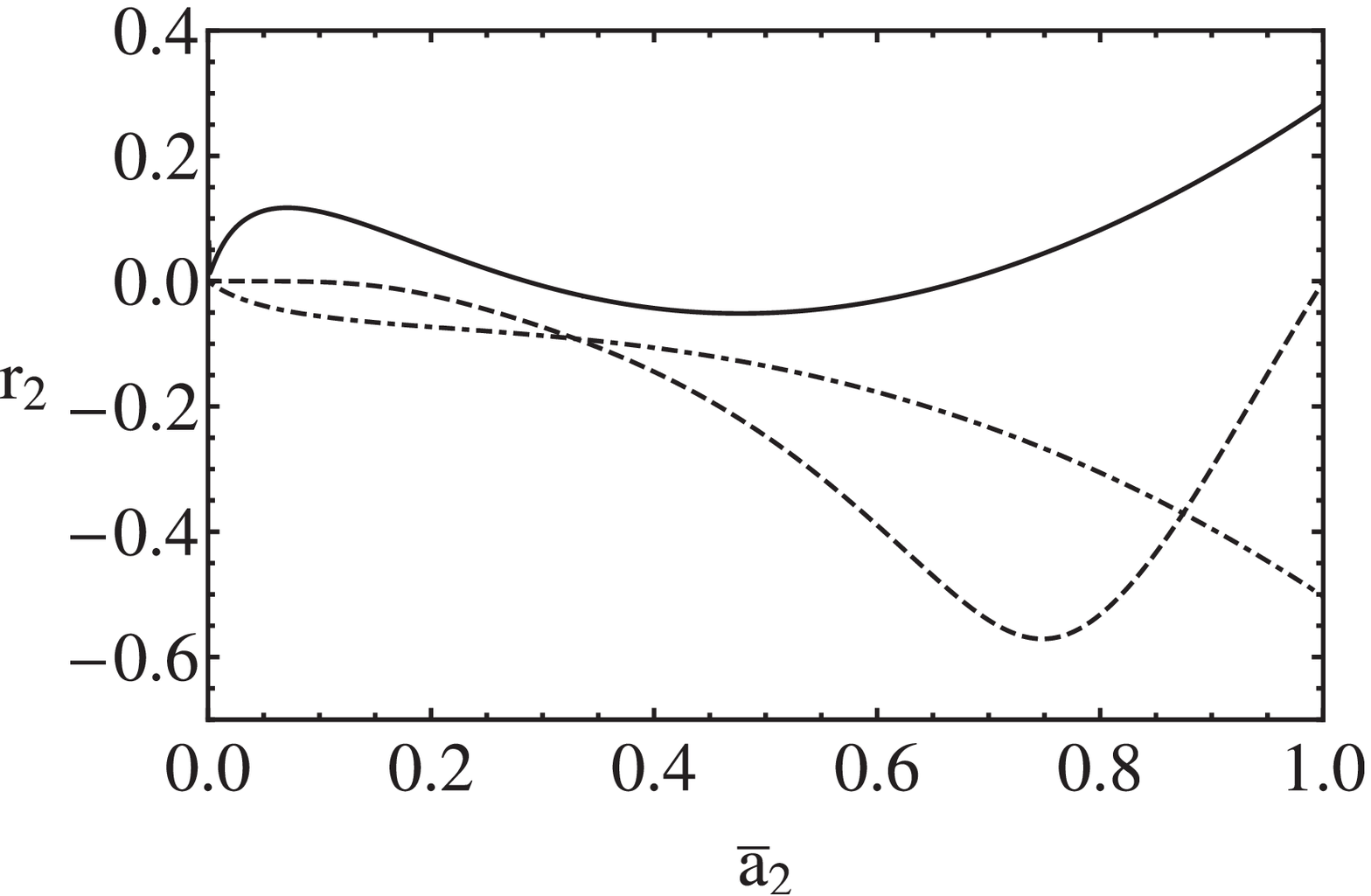}
\label{Fig:r_2}
} 
\subfigure{\includegraphics[scale=\scaleimagesS]{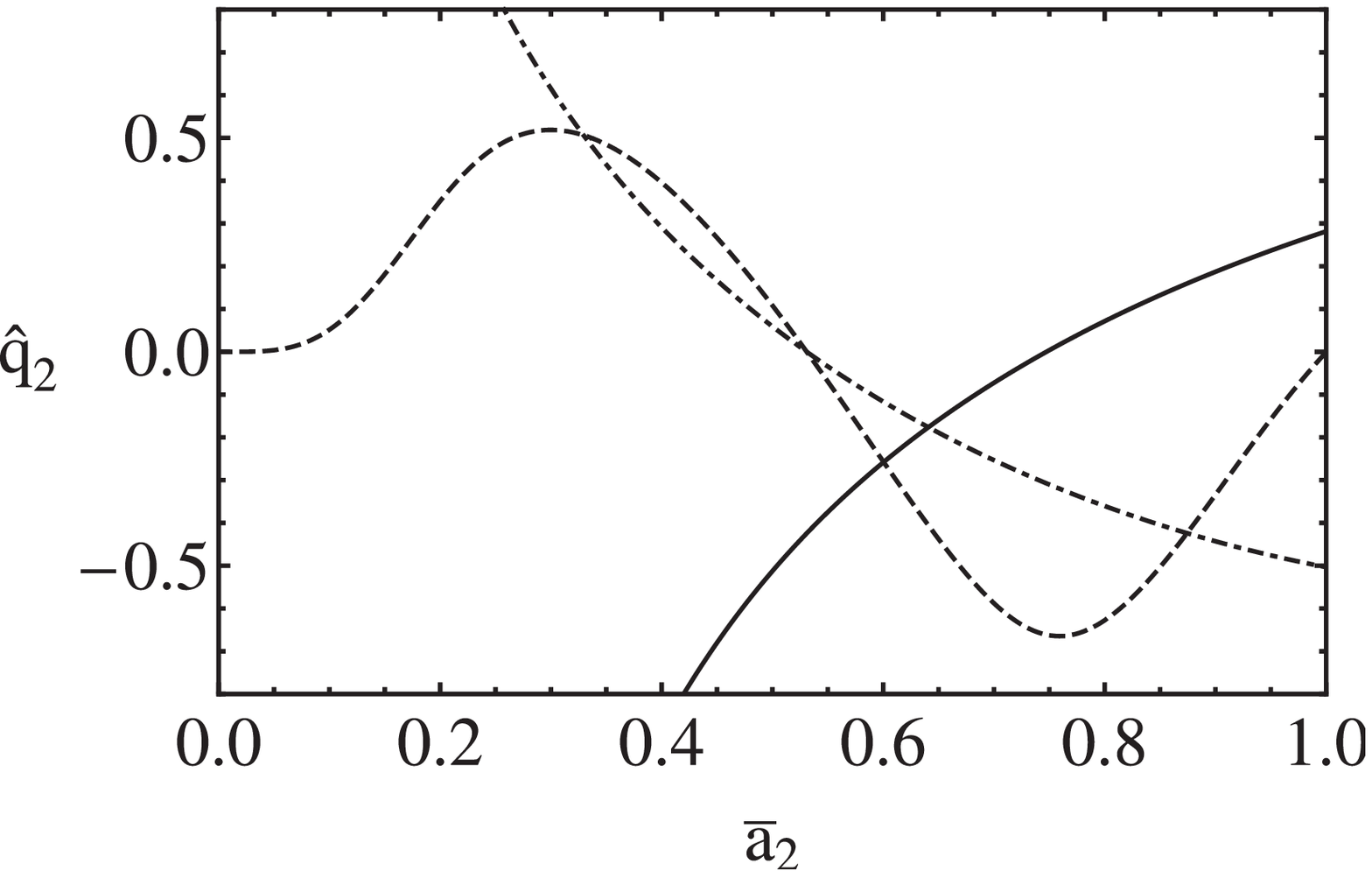}
\label{Fig:p_2}
}
\caption{{\it continued}}
\end{center}
\end{figure}

\newcommand{\scaleSurfacePN}{0.4}
\begin{figure}[htb!]
\begin{center}
\subfigure[$\hat w_1= 0.164537,~w_2=0.4$
$(w_1=0.4)$]{\includegraphics[scale=\scaleSurfacePN]{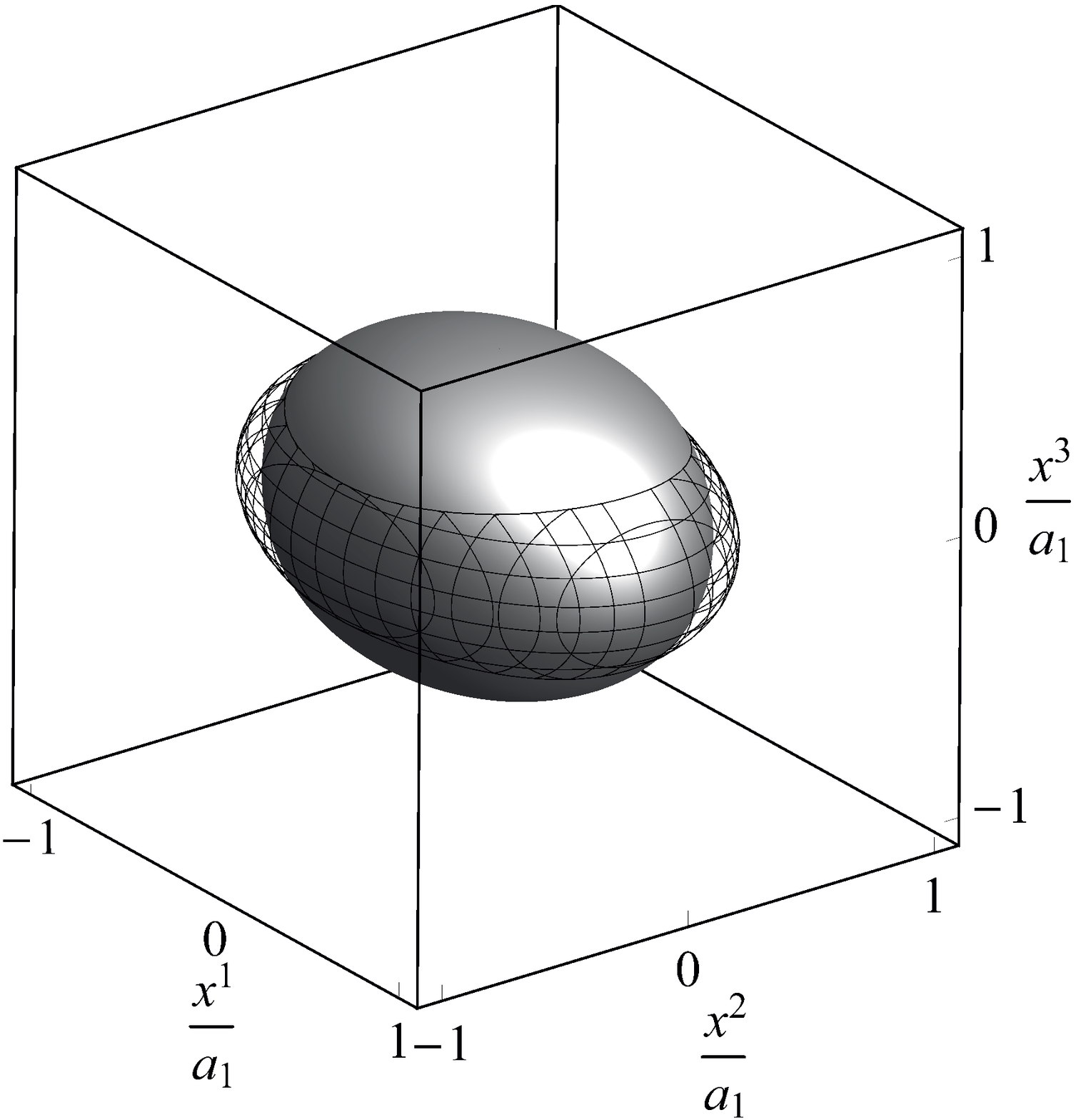}
\label{fig:surface_1}
}
\subfigure[$\hat w_1=0.468120,~w_2=0$ $(w_1=0)$, cf.
\cite{Chandrasekhar_197478}
]{\includegraphics[scale=\scaleSurfacePN]{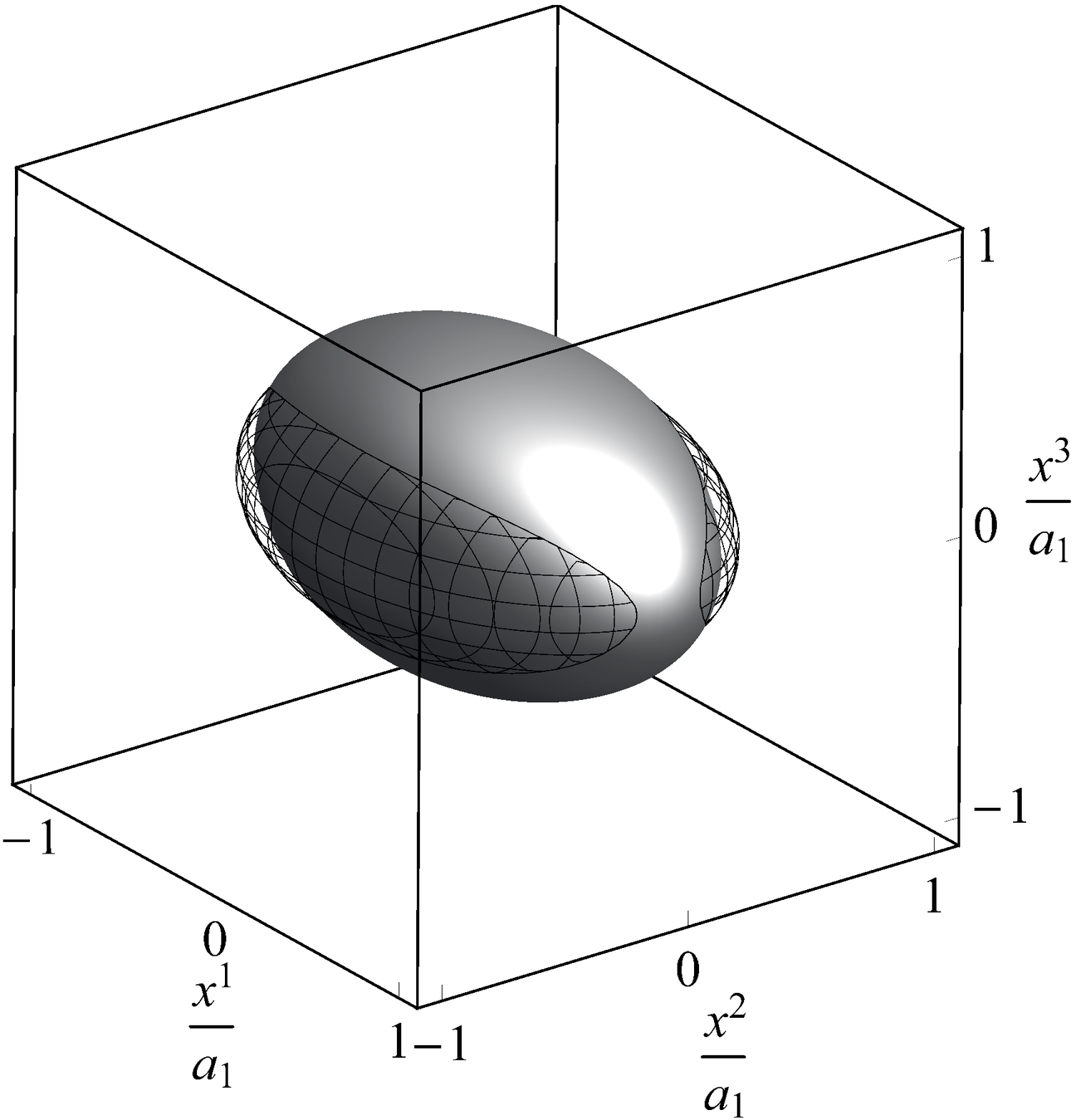}
\label{fig:surface_2}
}\\
\subfigure[$\hat w_1= -0.007157,~w_2=0.8$ $(w_1 =
0.7)$]{\includegraphics[scale=\scaleSurfacePN]{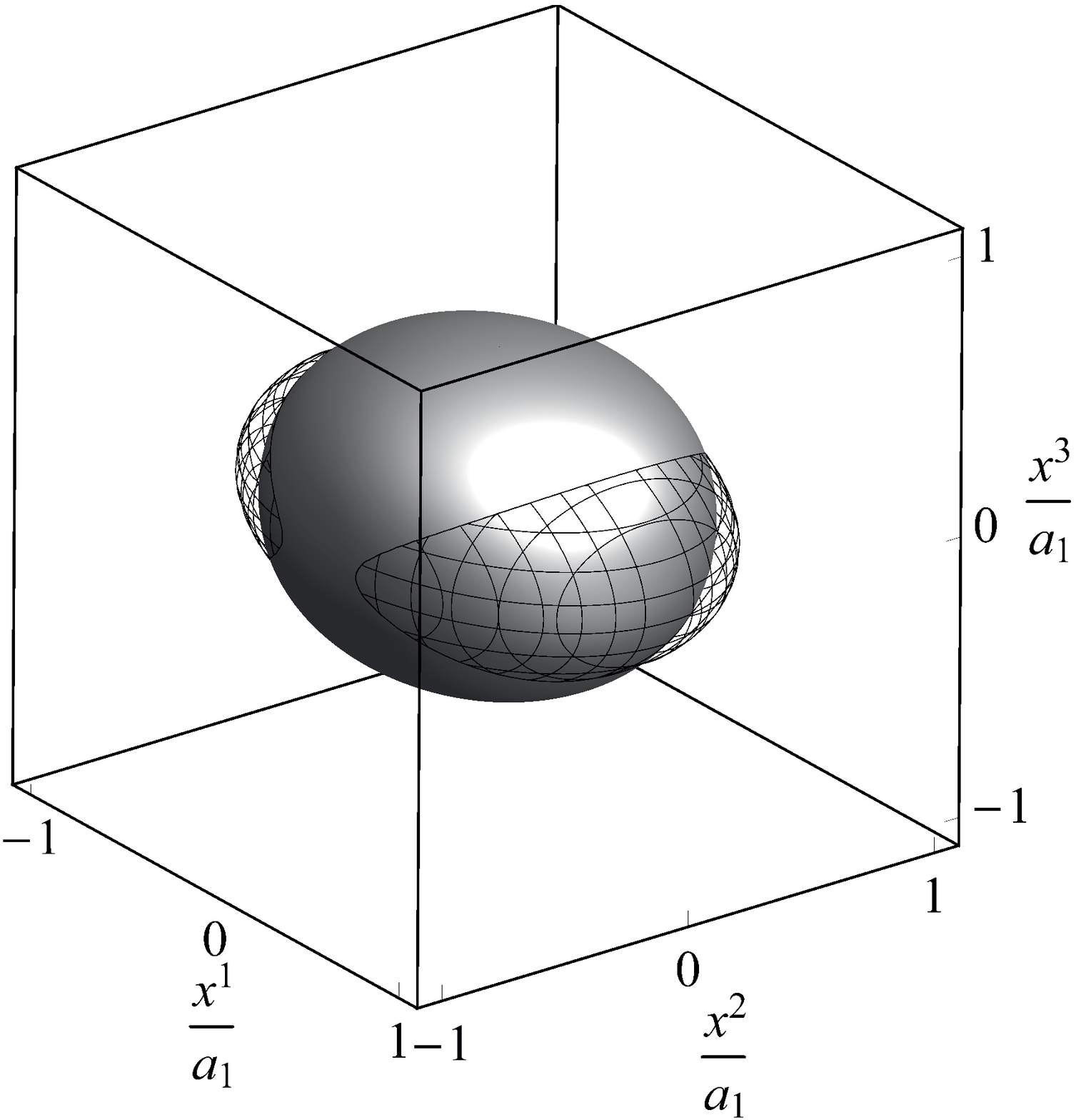}
\label{fig:surface_3}
}
\caption{\label{fig:surface} The 1 PN surfaces (solid, gray) compared to the
Newtonian surface (wireframe, black) for parameters $\ba_2=0.7$ and
$\tfrac{G\mu}{c^2}a_1^2=0.15$ and different choices of the $\hat w_1$ and
$w_2$.}
\end{center}
\end{figure}

\cleardoublepage
\end{document}